# 1I/'Oumuamua as an N₂ ice fragment of an exo-pluto surface II: Generation of N₂ ice fragments and the origin of 'Oumuamua

## S. J. Desch[1,*] and A. P. Jackson[1,^]

[1]School of Earth and Space Exploration, Arizona State University, Tempe, AZ 85287.

Corresponding author: Steve Desch ([steve.desch@asu.edu](mailto:steve.desch@asu.edu))

* Corresponding author, ORC-ID 0000-0002-1571-0836

^ ORC-ID 0000-0003-4393-9520

**Key Points:**

- A dynamical instability in the early Kuiper belt likely ejected ~$10^{14}$ N₂ ice fragments from thousands of plutos.

- A fragment of N₂ ice from an "exo-Pluto" would match all known attributes of the interstellar object 1I/'Oumuamua.

- The dynamical instability in the Kuiper belt experienced by our Solar System may be common among stellar systems.

## Abstract

The origin of the interstellar object 1I/'Oumuamua, has defied explanation. In a companion paper (Jackson & Desch, 2021), we show that a body of $N_2$ ice with axes 45 m × 44 m × 7.5 m at the time of observation would be consistent with its albedo, non-gravitational acceleration, and lack of observed CO or $CO_2$ or dust. Here we demonstrate that impacts on the surfaces of Pluto-like Kuiper belt objects (KBOs) would have generated and ejected $\sim 10^{14}$ collisional fragments— roughly half of them $H_2O$ ice fragments and half of them $N_2$ ice fragments—due to the dynamical instability that depleted the primordial Kuiper belt. We show consistency between these numbers and the frequency with which we would observe interstellar objects like 1I/'Oumuamua, and more comet-like objects like 2I/Borisov, if other stellar systems eject such objects with efficiency like that of the Sun; we infer that differentiated KBOs and dynamical instabilities that eject impact-generated fragments may be near-universal among extrasolar systems. Galactic cosmic rays would erode such fragments over 4.5 Gyr, so that fragments are a small fraction ($\sim 0.1\%$) of long-period Oort comets, but C/2016 R2 may be an example. We estimate 'Oumuamua was ejected about 0.4-0.5 Gyr ago, from a young ($\sim 10^8$ yr) stellar system, which we speculate was in the Perseus arm. Objects like 'Oumuamua may directly probe the surface compositions of a hitherto-unobserved type of exoplanet: "exo-plutos". 'Oumuamua may be the first sample of an exoplanet brought to us.

## Plain Language Summary

Our Kuiper belt originally had much more mass than today, but an instability caused by Neptune's migration disrupted their orbits, ejecting most of this material from the Solar System, and simultaneously causing numerous collisions among these bodies. There were thousands of bodies like Pluto, with $N_2$ ice (like the gas in Earth's atmosphere, but frozen) on their surfaces, and this instability would have generated trillions of $N_2$ ice fragments. A similar fragment, generated in another solar system, after travelling for about a half billion years through interstellar space, would match the size, shape, brightness, and dynamics of the interstellar object 1I/'Oumuamua. The odds of detecting such an object, as well as more comet-like objects like the interstellar object 2I/Borisov, are consistent with the numbers of such objects we expect in interstellar space if most stellar systems ejected comets and $N_2$ ice fragments with the same efficiency our solar system did. This implies other stellar systems also had Kuiper belts and similar instabilities. There are hints that some $N_2$ ice fragments may have survived in the Oort cloud of comets in our Solar System. 'Oumuamua may be the first sample of an exoplanet born around another star, brought to Earth.

# 1    Introduction

In October 2017, the object 1I/'Oumuamua was discovered by the Pan-STARRS telescope, soon after it passed only 0.22 AU from Earth and briefly reached 20[th] magnitude (Meech et al., 2017). Its orbit around the Sun was found to be hyperbolic, with eccentricity $e$=1.2, establishing 'Oumuamua as the first known object with a definitive origin outside the Solar System (Meech et al., 2017). A second such object, the interstellar comet 2I/Borisov, with eccentricity $e$=3.36, was discovered in August 2019[1]. These objects are just two of a population of millions of similar bodies that have been passing through the Solar System for centuries. They provide unique opportunities to probe the compositions and origins of materials ejected from extrasolar systems. In doing so, they provide a test of whether processes that occurred in our solar system are common in other planetary systems.

2I/Borisov is readily recognized as a comet similar to those in our solar system. It has actively outgassed species such as CN, and has a chemical composition similar to solar system comets that are depleted in carbon-chain species (Fitzsimmons et al., 2019; Opitom et al., 2019). Based on its trajectory, it may have been traveling at 23 km s[-1] at a distance of 14,000 AU from the M0v star Ross 573, 0.91 Myr ago This distance is consistent with an origin in the Oort cloud of Ross 573, even if the high ejection velocity is a challenge to explain (Bailer-Jones et al., 2020). Whatever details remain to be understood, 2I/Borisov shows every indication that it is comet-like, with a comet-like origin.

In contrast, 1I/'Oumuamua initially defied explanation. Bialy and Loeb (2018) listed several unusual properties of 'Oumuamua. It has been inferred—from calculations of the Pan-STARRS detection probability and the length of time the observatory has been operational—that each star in the Galaxy must have ejected ~$10^{15}$-$10^{16}$ such objects (Do et al., 2018), several orders of magnitude higher than predictions made prior to the discovery of 'Oumuamua (Moro-Martin et al., 2009). The velocity of 'Oumuamua with respect to the local standard of rest (LSR), the average velocity of stars in the Sun's neighborhood, was only 9 km s[-1]. This is far less than the typical tens of km s[-1] average dispersion of stars with respect to the LSR, and unexpected if 'Oumuamua were ejected from an average stellar system. 'Oumuamua's light curve indicated it was a highly elongated prolate or oblate ellipsoid, with axis ratios in the range of 5:1 to 10:1, more elongated than any known bodies in our solar system (Belton et al., 2018). The upper limits on its thermal emission (from *Spitzer Space Telescope* observations) indicated a body no more than a few hundred meters in size and an albedo unexpectedly high for asteroids or comets in the solar system (Trilling et al., 2018). Finally, from its trajectory, 'Oumuamua received a non-gravitational force outward, approximately proportional to $1/r^{1.5}$ or $1/r^2$ ($r$ is the distance from the Sun), with magnitude $4.9 \times 10^{-4}$ times the gravitational force (Micheli et al., 2018). This would be consistent with outgassing like comets experience, due to sublimation of ices on their sunward sides; but strict upper limits on the outgassing rates of $CO_2$ and CO, as well as dust, were imposed by *Spitzer Space Telescope* observations (Trilling et al., 2018). Despite suggestions that this non-gravitational force demanded properties like an artificial solar sail (Bialy and Loeb, 2018), 'Oumuamua is clearly a natural object; but it is unlike almost any other object in our solar system today.

---

[1] https://www.minorplanetcenter.net/mpec/K19/K19RA6.html

In a companion paper (Jackson & Desch, 2021), we demonstrate that many of the unusual properties of 'Oumuamua are explained if it is a large fragment of $N_2$ ice, with axes 45 m × 44 m × 7.5 m and mass $8.0 \times 10^6$ kg at the time it was observed at 1.42 AU. Its brightness would be consistent with an albedo of 0.64, which is exactly consistent with the albedo of the surface of Pluto, which is > 98% $N_2$ ice (Protopapa et al., 2017) and has a geometric albedo in the R band of $0.62 \pm 0.03$ (Buratti et al., 2015) and Bond albedo $0.75 \pm 0.07$ (Buratti et al., 2017). Jackson & Desch (2021) calculated the force associated with $N_2$ ice sublimating from 'Oumuamua's surface and calculated the resultant non-gravitational acceleration to be $4.9 \times 10^{-4}$ ($r$ / 1 au)$^{-1.8}$ cm s$^{-2}$, a perfect match to the observed values (Micheli et al., 2018). An object of fixed brightness has surface area that varies with albedo $p$ as $p^{-1}$, non-gravitational force due to sublimation that varies as $p^{-1}(1-p)$, and mass that varies as $p^{-3/2}$, so that its non-gravitational acceleration varies as $p^{1/2}(1-p)$. Because an 'Oumuamua with an albedo typical of $N_2$ ice, $p \approx 0.64$, is lower in mass, it can experience a high non-gravitational acceleration, as observed.

Moreover, a fragment of $N_2$ ice would suffer extreme mass loss as it passed perihelion at 0.255 au from the Sun. The calculations of Jackson & Desch (2021) demonstrate that 'Oumuamua would have decreased in mass from about $95 \times 10^6$ kg to $8 \times 10^6$ kg. The isotropic loss of material from a triaxial ellipsoid tends to increase axis ratios (Domokos et al., 2009), and Jackson & Desch (2021) calculate 'Oumuamua saw its axis ratios increase from 2.1:1 to 6.1:1 between entry into the Solar System and the time of the observations when it was at 1.42 au. This extreme mass loss, due in part to the volatility of $N_2$ ice, readily explains 'Oumuamua's uniquely large axis ratios.

Other unexplained aspects also are not as mysterious as they seemed at first, as the review by ('Oumuamua ISSI Team et al., 2019) makes clear. The low velocity of 'Oumuamua with respect to the LSR is not common among stars overall, but it is common among relatively young (< 2 Gyr-old) stellar system. Stars are born from molecular clouds, which have velocity dispersion ~6 km s$^{-1}$ with respect to the LSR, and acquire greater velocity dispersions over time through stellar encounters. As long as 'Oumuamua was ejected from a system during the first < 2 Gyr of its existence, it is actually likely that it would have such a low velocity dispersion, a point we return to in §3.1.

The relatively quick discovery of 'Oumuamua implies that interstellar objects are an order of magnitude more abundant than formerly thought (Jewitt, 2003). The detection of 'Oumuamua, combined with knowledge of current sky-survey detection limits, allows estimates of the density of objects in the ISM like 'Oumuamua. Do et al. (2018) estimated a density ~0.2 au$^{-3}$, while 'Oumuamua ISSI Team et al. (2019) estimated a number density ~0.1 au$^{-3}$, or ~$10^{15}$ pc$^{-3}$. Portegies Zwart et al. (2018), considering selection effects and statistics, calculated a probable range of density of objects, $3.5 \times 10^{13} - 2.1 \times 10^{15}$ pc$^{-3}$, with $7 \times 10^{14}$ pc$^{-3}$ considered likely. This is larger than had been expected. For example, up to ~$10^{13}$ comets are believed to have been ejected from the Solar System, based on the number inferred to exist in the Oort cloud. However, numerous models have been developed to explain how such a large number of objects could be produced and ejected from solar systems, summarized by 'Oumuamua ISSI Team et al. (2019).

Jackson & Desch (2021) estimated that 'Oumuamua's mass upon entry to the Solar System was $\sim 1 \times 10^8$ kg, and could have been $\sim 2.4 \times 10^8$ kg upon ejection from its Solar System, before passage through the interstellar medium (ISM). Using the latter mass, this implies a density of 'Oumuamua-like objects in the ISM of 0.0006–0.08 $M_E$ pc$^{-3}$. The masses involved are not great, and the discrepancy with the number of comets is not that large. Moreover, at only tens of meters in size, 'Oumuamua is much smaller than a typical comet, and our knowledge of the size frequency distribution of objects < 1 km has always been limited; as noted by ('Oumuamua ISSI Team et al., 2019), even $10^{15}$-$10^{16}$ total objects ejected per stellar system is not implausible.

Jackson & Desch (2021) concluded that 'Oumuamua was consistent with a fragment of $N_2$ ice produced by collisions with the surface of an exo-Pluto in another stellar system. A fragment tens of meters in size, made of $N_2$ ice with trace amounts of $CH_4$, would match 'Oumuamua's size, albedo, color, constraints on outgassed species, and especially the observed non-gravitational acceleration. The only unresolved question is whether it is possible to generate such large numbers—roughly $10^{15}$ per star—of fragments, tens of meters in size, from the surfaces of exo-Plutos in other stellar systems. A related question is whether a sufficiently large fraction of the fragments ejected from typical stellar systems are $N_2$ ice in particular, so that it is not improbable that one would be the first interstellar object detected.

In this paper we test the hypothesis that 'Oumuamua was a fragment of $N_2$ ice generated by collisions with the surface of a Pluto-like body in another stellar system. In §2 we calculate how many $N_2$ ice and other fragments of Pluto-like bodies must have been ejected from theSolar System, and extrapolate to derive statistics about the number and fraction of interstellar objects that are similar fragments of exo-Plutos. In §3 we consider implications for the probability of detecting objects like 1I/'Oumuamua and 2I/Borisov, the possibility of detecting $N_2$ ice fragments among long-period comets, and the universality of the processes in the Solar System that could have led to ejection of $N_2$ ice fragments. We summarize our findings in §4 and conclude that 'Oumuamua is very plausibly a fragment of an exo-Pluto, and that the geophysical processes and dynamical instabilities that led to such bodies being ejected from our Solar System must be somewhat universal among stellar systems.

## 2 Generation of fragments during the dynamical instability in the Kuiper Belt

### 2.1 Total mass of KBO surface fragments lost from the Kuiper Belt

We hypothesize here and in the companion paper (Jackson & Desch, 2021) that 'Oumuamua was originally a chunk of $N_2$ ice with mass $\sim 2.4 \times 10^8$ kg and mean radius $\sim 40$ m, generated by collisions on the surface of a Pluto-like body and then ejected from another stellar system. To judge whether this is a plausible scenario, we first calculate the likelihood that such a body would be generated and ejected from our own Solar System. We first focus on the specific questions of how much mass of fragments would be generated from collisions during the dynamical instability that depleted the Kuiper Belt, then ask what their typical sizes, numbers, and compositions would be.

The opportunity to generate collisional fragments and eject them from the solar system would have occurred during the motion of Neptune through the primordial Kuiper Belt (Malhotra, 1993). As predicted by the Nice model (Tsiganis et al., 2005), almost all of the 35 Earth masses of material in the 15-30 AU region was ejected (mostly by Jupiter), either to the Oort cloud or interstellar space. More recent models (Nesvorný and Vokrouhlický, 2016) favor slightly lower initial masses, ~ 20 $M_E$. The present-day mass of the Kuiper Belt is estimated as ~0.02 $M_E$ (Pitjeva and Pitjev, 2018), so all but a fraction ~$10^{-3}$ of the primordial Kuiper Belt was ejected. The mass ejected from the outer Solar System far surpasses the mass ejected from the asteroid belt, which was no more than a few Earth masses of material (Shannon et al., 2015, 2019), possibly less (Morbidelli and Raymond, 2016). Ejection of collisional fragments along with the larger bodies is equally likely.

Collision of KBOs with Pluto-like bodies is a likely event. A significant fraction of the primordial Kuiper belt mass was in the form of Pluto-sized (radius ~1188 km, mass ~0.0022 $M_E$) or Triton-sized (radius ~1353 km, mass ~ 0.0036 $M_E$) objects. Nesvorný and Vokrouhlický (2016) demonstrated that the distribution of Kuiper Belt Objects (KBOs) between resonant and non-resonant (with Neptune) objects is best explained by having Neptune interact with roughly 2-8 $M_E$ of such large objects, representing 10-40% the mass of the Kuiper belt they considered. They tested a few scenarios: one with either 1000, 2000, or 4000 Pluto-sized objects; one with 1000 objects twice the mass of Pluto; and one with a mix of ~1000 Plutos and ~500 two-Pluto-mass objects. Because the interaction of Neptune with large objects introduces a "graininess" to its migration, all these models were successful in reproducing the distribution of resonant vs. non-resonant objects. It is sensible that if the Kuiper belt was depleted by a factor of ~$10^3$ in mass, that the presence of a few large dwarf planets (Pluto, Eris, Triton) would have demanded ~3000 Pluto-sized objects initially, totaling about 6 Earth masses.

The existence of thousands of Pluto-sized objects may be surprising, but is broadly supported by a large number of models for how bodies grow in the Kuiper belt, as reviewed by Shannon and Dawson (2018). The "pebble-pile" model of Hopkins (2016) generally predicts ~$10^3$ Pluto-sized objects, and about 3 times as many objects the size of the KBO Gonggong (2007 $OR_{10}$), which has a mean radius 615 km and mass 0.00029 Earth masses (Kiss et al., 2019). Hopkins (2016) assumed a very low-mass disk, with mass 0.008 $M_\odot$, smaller than even traditional minimum-mass solar nebula models (Weidenschilling, 1977; Hayashi, 1981). As discussed by Desch (2007), who updated the minimum-mass solar nebula model to account for the 35 Earth masses of objects in the primordial Kuiper belt as in the Nice model (Tsiganis et al., 2005), the likely mass is at least 4, possibly 8 times larger than this. This suggests that the numbers of large KBOs are nearer the upper limits allowed by the analysis of Shannon and Dawson (2018): about 2000 plutos and 6000 gonggongs, adding up to about 6 $M_E$. This mass is in the range constrained by Nesvorný and Vokrouhlický (2016). This represents a not unreasonably high fraction of the total mass of the primordial Kuiper belt, which was at least 20 $M_E$ (Nesvorný and Vokrouhlický, 2016). but perhaps up to 35 $M_E$ (Tsiganis et al., 2005). If the primordial Kuiper belt had 35 Earth masses, it has been depleted by a factor of about 1800, leaving about 3 gonggongs and 1 plutos, roughly consistent with the presence of about 4 Gonggong-sized objects (Gonggong, Quaoar, Makemake, Haumea) and 2 Pluto-sized objects (Pluto and Eris, although Triton should be considered). For concreteness, we will assume that the primordial Kuiper belt held 2000 objects ('plutos') with radius ~1200 km, and 6000 objects ('gonggongs') with radius ~600 km, each with density ~1.8 g cm$^{-3}$, totaling about 6 Earth masses.

The rest of the mass of the primordial Kuiper belt would be in the form of small KBOs, mostly with diameters $D \sim 100$ km, but also including fragments from their mutual collisions. The size distribution of KBOs in the dynamically excited Kuiper belt from observations of KBOs is $dN/dD \propto D^{-q}$, with $q = 2^{+0.5}_{1}$, for objects with $D < 110$ km, and $q = 5.3^{+0.4}_{1}$ for larger objects (Fraser et al., 2014). This compares with their values for the dynamically cold population of $q = 2.9 \pm 0.3$ for objects with $D < 140$ km, and $q = 8.2 \pm 1.5$ for larger objects. An alternative estimate of the size distribution at small sizes comes from Singer et al. (2019), who examined small craters on Pluto and Charon, suggesting that $q \sim 1.7$, though the effects of crater degradation are difficult to quantify. Presumably KBOs are formed as objects with sizes ~110 km or 140 km, and smaller objects represent a collisional cascade as in the asteroid belt, although a slope of $q \approx 2 - 2.9$ is somewhat shallower than would usually be expected from collisional equilibrium, which would typically yield values of $q$ in the range 3-4 depending on how the strength of the bodies varies with size. The KBOs with very small sizes may be associated with comet nuclei. Because it has not been excited or suffered as many collisions, the dynamically cold KBOs may represent a more primordial population. For concreteness we adopt the dynamically cold population, with $q = 2.9$ for 1 km $< D < 140$ km and $q = 8.2$ for $D > 140$ km. For $D < 1$ km we assume a power law of $q = 1.7$. We take the total mass to be 29 M$_E$. For an internal density 1 g cm$^{-3}$, this yields $6.7 \times 10^{11}$ KBOs/comets with $D > 1$ km, and $1.3 \times 10^{13}$ with $D > 50$ m.

Following the dynamical instability associated with triggering Neptune's migration and the orbital excitation of the Kuiper Belt, the population of small KBOs would have collided not just with each other, but with the thousands of plutos and gonggongs, generating a third population of objects: the collisional fragments from the surfaces of differentiated KBOs. We can estimate the erosion of each pluto's surface due to this process, constraining the total mass that collided, and the velocity. The collisional grinding of the Kuiper belt appears attributable to objects that are now part of the scattered disk population (Morbidelli and Rickman, 2015), so we assume that the colliding KBOs in the primordial Kuiper belt between 15 and 30 AU were dynamically excited to have average eccentricites $e \sim 0.4$ and inclinations $i \sim 20°$, consistent with the scattered disk. Their velocities relative to the large KBOs then would have been $V_{rel} \sim v_K (e^2 + i^2)^{1/2} \sim 3.5$ km s$^{-1}$ (at 20 au).

The mass ejected from each body scales with the mass impacting each body, and depends on the ratio of the impact velocity to the escape velocity. The escape velocity from a gonggong is $V_{esc} \approx 0.60$ km s$^{-1}$, and from a pluto is 1.2 km s$^{-1}$. KBOs colliding with heliocentric relative velocity 3.5 km s$^{-1}$ would have impact velocities $V_{imp} \approx 3.6$ km s$^{-1}$ and 3.7 km s$^{-1}$. The mass of material ejected from an icy body by impactor with mass $M_{imp}$ and striking at an angle $\theta$ from the normal is given by $M_{imp} C (V_{esc} / V_{imp} \sin \theta)^{-3\mu}$, where $\mu = 0.55$ and $C = (3k/4\pi)C_0^{3\mu} = 0.14$, where $k = 0.3$ and $C_0 = 1.5$ (Housen and Holsapple, 2011; Hyodo and Genda, 2020). Thus we predict a head-on impactor at a typical speed would generate $F_g = 2.6$ times its mass in debris from each gonggong, and $F_p = 0.9$ times its own mass in debris from each pluto. Accounting for the fact that not all collisions are head-on would lower the mass of debris, but accounting for the effect of gravitational focusing on the collision cross section and the distribution of impact angles increases the mass of the debris, and the two factors largely cancel each other out.

The mass of KBO impactors actually colliding with the plutos and gonggongs is a fraction of the total mass of smaller KBOs. We assume the instability of Neptune's outward migration, and the depletion of the Kuiper belt, took an $e$-folding time $t \sim 50$ Myr (Malhota, 1993; Tsiganis et al., 2005; Nesvorný and Vokrouhlický, 2016), and that the large KBOs are distributed in volume $1.7 \times 10^4$ au$^3$ between $d$=15 and 30 au, with thickness $\sim d / 3$ (based on excitation to inclinations $i\sim20°$). The number density and cross section of gonggongs would be $n_g \sim 0.36$ au$^{-3}$ and $\sigma_g = \pi(600$ km$)^2$, and that of plutos would be $n_p \sim 0.12$ au$^{-3}$ and $\sigma_p = \pi(1200$ km$)^2$. Moving through these bodies at a relative velocity $V_{rel} \sim$3.5 km s$^{-1}$ results in a collision probability $[n_g \sigma_g + n_p \sigma_p] V_{rel} t \sim 1.6 \times 10^{-3}$. This is a negligible fraction of the 29 Earth masses of the smaller KBO population, but still would generate a considerable amount of debris from the surfaces of large KBOs: $M_{imp} [n_g \sigma_g F_g + n_p \sigma_p F_p] V_{rel} t \sim 0.074$ M$_E$.

During the dynamical instability that depleted the Kuiper Belt, the amount ejected per pluto would be about $1.2 \times 10^{-5}$ M$_E$, or 0.5% of its mass, equivalent to erosion of a global layer about 4 km thick. The amount ejected per gonggong would be about $8.4 \times 10^{-6}$ M$_E$, or 3.1% of its mass, equivalent to a global layer about 11 km thick.

The vast majority of impacting objects would have been smaller than around 100 km in diameter (below the knee in the size distribution found by Fraser et al. (2014), but each Pluto may be struck by a few larger objects. In a distribution with $q$=2.9 and 8.2 above and below a break at 140 km, like that discussed above, with a total mass of 29 Earth masses, we estimate the number of KBOs (with presumed density 1.8 g cm$^{-3}$) with diameters > 10 km would be $N_{KBO} \sim 8.4 \times 10^9$, the number with $D > 100$ km would be $\sim 6.5 \times 10^7$, and the number with $D > 300$ km would be $\sim$6.1 x $10^4$. These estimates are consistent with those presented by Shannon and Dawson (2018). The number of KBOs (with typical inclination $i\sim20°$) with $D > 10$ km a given pluto would collide with would be $[ N_{KBO} / (1.7 \times 10^4$ au$^3)] \times [\pi (1200$ km$)^2 ] \times [3.5$ km s$^{-1}] \times [50$ Myr$] \approx$ 3700. The impactor that produced Sputnik Planitia on Pluto is believed to have had $D \approx$ 150-300 km (McKinnon et al., 2017). The number of very large KBOs with $D > 150$ km (or $D > 300$ km) a pluto would collide with would be $\approx 4$ (or 0.03), consistent with the existence of a single large impact basin on Pluto if it requires a $\sim$180 km impactor. Overall, though, very few objects would produce craters this deep, and it is unlikely that a gonggong would be struck by even one object > 150 km.

For the most part, each pluto is "sand-blasted" to an average depth of only $\sim$4 km, and each gonggong to an average depth of only $\sim$11 km, by predominantly much smaller KBOs. We conclude that about 1.2% of the combined mass of all $\sim$2000 plutos and $\sim$6000 gonggongs (totaling 6 M$_E$) in the Kuiper Belt was ejected from these large bodies. This population of $\sim$0.074 Earth masses of surface fragments would match the composition of the uppermost few km of each object, which we now argue would have been N$_2$ ice.

## 2.2 Composition of the collisional fragments from differentiated KBOs

Many fragments from the surface of each large differentiated KBO very likely were composed of $N_2$ ice. $N_2$ ice covers the surface of Triton to depths > 1 km (Cruikshank et al., 1995), and covers 98% of Pluto's surface today, with frosts of $CH_4$ and CO making up the remainder (Owen et al., 1993; Protopapa et al., 2017). It is inferred to fill the Sputnik Planitia basin, which is still 2-3 km deep, possibly up to 10 km, and convection cell patterns in the $N_2$ ice require depths of several km (McKinnon et al., 2016). The amount of ice in Sputnik Planitia is equivalent to a global layer 200-300 m thick (McKinnon et al., 2016), but could have been much larger in the past. Similarly, Triton's surface today is dominated by global layer of $N_2$ ice estimated to be about 1-2 km thick (Cruikshank et al., 1998). It is not precisely known how deep the $N_2$ ice is on Pluto or Triton, or how much has been lost over Solar System history.

A more direct estimate may come instead from an accounting of the nitrogen inventories within large KBOs. The cosmic abundance of nitrogen (Lodders, 2003) allows for $N_2$ ice to be as high as 16% the mass of $H_2O$ ice. This fraction of the ice layer, which is about 1/3 of Pluto's mass, suggests an original mass of $N_2$ ice of 0.05 Pluto masses, forming a layer about 35 km thick, on top of $H_2O$ ice. On a Gonggong-like body with radius 600 km, the thickness would be about 18 km. These thicknesses far exceed the depths to which plutos were eroded by collisions (~4 km), and even the depths to which gonggongs were eroded (~11 km), suggesting ejection of pure $N_2$ ice fragments. The requirements for such a thick layer and for the ejected fragments to mostly be $N_2$ ice are: 1) KBOs accrete the cosmic abundance of N; 2) the N is converted efficiently to $N_2$; 3) the $N_2$ is transported to the surface; 4) this occurs before the erosion accompanying the dynamical instability; and 5) the surface $N_2$ ice does not sublimate before the dynamical instability.

These conditions are probably met in KBOs the size of Gonggong and Pluto. The first requirement probably would not be met if $N_2$ had to condense directly from the solar nebula. $N_2$ would require temperatures < 40 K to condense, and comets are notoriously depleted in $N_2$, leading to identification of the "missing nitrogen" problem (Poch et al., 2020). This problem may have been resolved by near-infrared (NIR) spectroscopy of comet 67P/Churyumov-Gerasimenko that has identified ammonium salts as a major reservoir of N, about half the cosmic abundance, the other half residing in macromolecular organics. This implies that only a few percent of the cosmic abundance of nitrogen was in the form of volatile $N_2$ (Poch et al., 2020). Thus comets, and presumably large KBOs, could have accreted almost the full cosmic abundance of nitrogen in solid form.

The second requirement is that the nitrogen is efficiently converted into $N_2$. This requires chemistry within a subsurface ocean with relatively high temperatures and/or oxidizing conditions (low hydrogen fugacity). As reviewed by McKinnon et al. (2020), nitrogen—whether in organic material, ammonia, or ammoniated minerals—is efficiently converted into $N_2$ in subsurface oceans or, more likely, hydrothermal circulation through a rocky core. Models predict that substantial subsurface oceans may exist on Pluto for a wide range of input parameters, as long as the outermost ice shell is not convective (Robuchon and Nimmo, 2011). Subsurface oceans are produced even on bodies as small as Charon and Gonggong with radii ~600 km (Desch et al., 2009). Although $N_2$ ice is difficult to observe spectrally, shifts in the wavelengths at which $CH_4$ absorbs sunlight strongly suggest $N_2/CH_4$ ices exist on the surfaces of Eris, Quaoar, and Makemake, with $N_2$ possibly dominating (Tegler et al., 2008, 2010; Barucci et al., 2015; Lorenzi et al., 2015). There would seem to be ample opportunity for ocean chemistry to convert nitrogen to $N_2$, as has apparently happened on other KBOs.

The third requirement is that the $N_2$ is transported to the surface. This may be effected through cryovolcanism (Neveu et al., 2015). $N_2$ is relatively insoluble in liquid and is not expected to reside in clathrate hydrates (Kimura and Kamata, 2020). The fact that ample $N_2$ resides on the surfaces of Triton, Pluto, and Titan, strongly suggests $N_2$ can fully outgas under a variety of planetary conditions. Because other species are more soluble in oceans or taken up in clathrate hydrates (Kimura and Kamata, 2020), the surfaces of KBOs can be expected to be largely pure $N_2$ ice.

The fourth requirement is that the geochemical evolution of KBOs occurs before the dynamical instability that erodes their surfaces and ejects material. For our solar system, the first models suggested that would take place at 650 Myr, coincident with the Late Heavy Bombardment (Tsiganis et al., 2005). The imperative for this timing has been removed, and more recent studies favor an early outward migration of Neptune at < 100 Myr, to account for binaries within the Trojan asteroids (Nesvorný et al., 2018) and to explain other dynamical aspects of the Kuiper belt (de Sousa et al., 2020). Other models (Clement et al., 2018) suggest Neptune migrated perhaps closer to 10 Myr after the birth of the Solar System. Models overwhelmingly show that extensive oceans can form on Charon/Gonggong-sized bodies and Pluto-sized bodies on timescales < 100 Myr (Desch et al., 2009; Robuchon and Nimmo, 2011; Kimura and Kamata, 2020; Canup et al., 2020). According to the "hot early start" model for Pluto, it must have begun its geophysical evolution with a significant subsurface ocean (in < $10^5$ years), or else the initial melting of ice to form the ocean would have manifested itself as compressional tectonic features on Pluto's surface (Bierson et al., 2020). That suggests differentiation and production of $N_2$ ice within < 10 Myr. As long as a system is not marked by both late (~100 Myr) differentiation of KBOs and early (~10 Myr) migration of Neptune, the instability is likely to take place before the $N_2$ ice crusts are formed.

Finally, the fifth requirement is that large KBOs could retain $N_2$ ice on their surfaces for ~$10^8$ yr. Even at temperatures > 50 K, Pluto-sized KBOs are capable of retaining $N_2$ ice on their surfaces for the age of the Solar System, but Gonggong-sized KBOs would require $T$ < 34 K to do so (Schaller and Brown, 2007). We calculate the Jeans mass loss rate to be roughly 20 times faster at 40 K, and 500 times faster at 50 K, so that $N_2$ ice surfaces could be retained by gonggongs for only ~$2 \times 10^8$ yr and ~$10^7$ yr at these temperatures. Assuming $p_B \approx 0.85$ and ε~0.85, like on Triton, and the Sun's luminosity was 0.7 $L_\odot$ at this early time, the temperature would have been < 43 K beyond 15 AU. This means retention of $N_2$ ice on the surfaces of gonggongs was likely, and retention of $N_2$ ice on the surfaces of plutos was ensured, for ~$1 \times 10^8$ yr, throughout the primordial Kuiper belt.

Combined, these considerations make it likely that a very large percentage, perhaps all, of the cosmic abundance of nitrogen could be converted into $N_2$ by chemistry in the subsurface oceans, then outgassed to form a thick layer of nearly pure $N_2$ ice on the surface up to 35 km thick on each pluto, or 18 km thick on each gonggong. Below this, presumably, would be layers of $H_2O$ ice. This would occur very early in each large KBO's evolution, and probably before the onset of the dynamical instability. Erosion of the topmost few km would generate fragments that are almost entirely $N_2$ ice, possibly with small amounts of $CH_4$ and CO ice dissolved in it.

While this reasoning makes it likely that the dynamical instability in the Kuiper belt generated many $N_2$ ice fragments, it is necessary to show that it would not also generate a much larger number of fragments of other composition, especially $H_2O$ from the underlying 'bedrock' on large KBOs; otherwise, the existence of 'Oumuamua as an $N_2$ ice fragment would still be improbable. Most impactors are considerably smaller in diameter than the tens of km thicknesses of the $N_2$ ice shell we propose, and would only eject $N_2$, but most of the impacting mass is in larger impactors. Since the ejecta mass scales with this mass, most of the ejecta will be generated by impactors on the larger size. Based on the size distribution above (§2.1), the largest impactor likely to hit a pluto has diameter ~180 km, and the largest impactor likely to hit a gonggong has diameter ~150 km. Crater scaling relations for $N_2$ ice surfaces are not available, but using the values for $H_2O$ ice suggested by Kraus et al. (2011) indicates the transient crater should be roughly 5 times the diameter of the impactor. This will subsequently collapse into a wider and shallower complex crater, but only the excavation phase is relevant. The depth of the transient crater is ~1/3 of its diameter; however excavated material only originates from the upper parts of the transient crater, down to ~1/3-1/2 its depth (Melosh, 1989). As such, we expect that for an impactor of diameter $D$, the maximum depth of excavated material is ~0.6 $D$. We assume that on a gonggong, if 0.6 $D$ < 18 km (i.e., $D$ < 30 km), all the ejected material is $N_2$ ice; while if $D$ > 30 km, the fraction of ejected material that is $N_2$ ice is (30 km / $D$). For plutos, with a 35 km thick $N_2$ ice layer, the fraction is (60 km / $D$). Integrating over the distribution of impactor masses, we calculate that roughly 45% of the mass ejected from gonggongs is $N_2$ ice, as is 69% of the mass ejected from plutos. In all, of the 0.074 $M_E$ of fragments ejected from the surfaces of these large KBOs, about 0.043 $M_E$, or about 60% of the excavated material, will be $N_2$ ice fragments, with somewhat more $N_2$ coming from plutos than gonggongs.

## 2.3    Numbers and masses of the $N_2$ ice surface fragments generated

We predict that a total mass $\approx$ 0.074 $M_E$ of fragments from the surfaces of large KBOs was generated during the dynamical instability in the Kuiper belt. Of this, $\approx$ 0.043 $M_E$ were $N_2$ ice fragments. Below we argue that about 80% of all surviving fragments (i.e., 0.013 $M_E$), were ejected from the solar system. This mass compares not unfavorably with the masses we inferred must be ejected per 1 $M_\odot$ of star, ~ 0.23 $M_E$ . Our estimate is lower than the best guess for the number density of fragments by Portegies Zwart et al. (2018) and Widmark and Monari (2019), at their lower limit, ~0.012 $M_E$. However, to check whether the number of ejected fragments matches the inferred ~$10^{15}$ fragments like 'Oumuamua that must be ejected per stellar system, we must quantify the average mass of fragments.

The average mass of collisional fragments depends on their size distribution. Studies of secondary craters on Mars (Robbins and Hynek, 2011) suggest fragments from a single impact obey a differential size distribution with a slope that is on average around $q\approx$5, albeit with a large possible range from 3.3 to 8, up to a maximum fragment diameter ~0.1 × that of the impactor. Combined with the $q\approx$2.9 distribution that we might expect for the impactors (below ~140 km) as described above, this would potentially suggest a remarkably steep profile like $q$ > 6 for the $N_2$ ice fragments, absent any further collisional processing within the fragment distribution. In such a distribution, the average value of $D^3$ is $(q-1)/(q-4) D_{min}^3 = 2.5 D_{min}^3$ for $q$=6, where $D_{min}$ is the diameter of the smallest fragments. The value of $D_{min}$ is uncertain, but given that 'Oumuamua itself was eroded by tens of m within the Solar System, and possibly ~10 m while in the ISM, we consider $D_{min}$~50 m to be reasonable, or else the typical fragment would not survive to be observed. The typical diameter would then be ~70 m and the average mass of a fragment would then be $2.5 \times (\pi/6)\rho D_{min}^3$, which for $D_{min}$=50 m and density $\rho$ ~1 g cm$^{-3}$ is ~$1.6 \times 10^8$ kg, matching our estimate for 'Oumuamua's original mass. With this average mass, the number of fragments in the early Solar System would have been ≈ $(0.074 \, M_E) / (1.6 \times 10^8 \, kg) = 2.7 \times 10^{15}$. This refers to the number of fragments (both $H_2O$ and $N_2$ ice) with $D > 50$ m. The cumulative number with $D > 1$ km would have been a factor $(1 \, km \, / \, 50 \, m)^{-5}$ lower, only ~$8 \times 10^8$.

This number also must be compared to the number of KBOs with $D > 50$ m, to ensure that an even larger number of such objects are not also ejected from the Solar System. Unfortunately, this number is difficult to predict, as it is unknown how to extrapolate the KBO size distribution to sizes smaller than ~10 km. For $D > 10$ km, there are about $10^8$ such objects (Shannon and Dawson, 2018). Assuming a size distribution for the dynamically cold distribution, d$N$/d$D \sim D^{-q}$, with $q \approx 2.9$ below 140 km (Fraser et al., 2014) implies about $6.7 \times 10^{11}$ KBOs with $D > 1$ km, and $1.3 \times 10^{13}$ KBOs with $D > 50$ m. Thus KBOs would represent a small fraction (<1%) of the number of objects > 50 m, and would not significantly change the numbers we derived above. KBOs would, however, dominate objects > 1 km in size.

A different approach to estimating the numbers of small KBOs (comets) ejected from the solar system is to determine the number trapped in the Oort cloud, and the efficiency by which they are trapped in the Oort cloud vs. ejected from the solar system. Objects residing in the Oort cloud are far too faint to be directly observed (although occultations can be used to provide constraints (Chang et al., 2016). Instead, the number of objects in the Oort cloud must be extrapolated from the number of long-period comets that derive from it. Boe et al. (2019) use data from the Pan-STARRS1 survey to examine the magnitude distribution of cometary nuclei, finding that they obey a distribution $dN/dH_N \propto 10^{\alpha H_N}$, where $H_N$ is the absolute H-band magnitude of the nucleus; they find that α=0.7 for $H < 16.9$ and α=0.07 for $H > 16.9$. Assuming an albedo of 0.04 that is independent of size, this corresponds to a size distribution $dN/dD \propto D^{-q}$, where $q$=4.5 for $D > 2.8$ km and $q$=1.35 for $D < 2.8$ km. This leads to a total mass for the outer Oort cloud of around 1.3 $M_E$ and around $1.5 \times 10^{12}$ objects with $D > 1$ km, and $4.3 \times 10^{12}$ objects with $D > 50$ m, although bodies less than around 3 km in diameter contribute little to the total mass, due to the much shallower slope below this size. This mass estimate for the outer Oort cloud is comparable to those by previous authors, as discussed by Boe et al. (2019) and Dones et al. (2015).

During the dynamical instability that ejected them, comets were trapped in the outer Oort cloud with low efficiency, ~1-2%, but a dynamical pathway for populating long-period comets from the inner Oort cloud has been established, and the inner Oort cloud can trap 5-10 times more of ejected planetesimals if the dynamical instability took place while the Sun still resided in a stellar cluster (Brasser et al., 2006; Kaib and Quinn, 2008, 2009). The mass of the inner Oort cloud is estimated to be up to 5 times as large as that of the outer Oort cloud (Dones et al., 2004, 2015). Estimates of the fraction of all objects trapped in the Oort cloud (instead of ejected from the Solar System) range from 5-7.6% (Dones et al., 2004 to ~5% (Vokrouhlický et al., 2019) to ~7% (Brasser and Morbidelli, 2013) at the low end, to values more like 10-15%, with a maximum of 20% at the high end (Shannon and Dawson, 2018). In the combined inner and outer Oort cloud there could be perhaps 7.8 $M_E$. and around $9 \times 10^{12}$ objects with $D$ >1 km, and $2.6 \times 10^{13}$ objects with $D$ > 50 m. To be conservative about the mass ejected from the Solar System, we assume the efficiency of trapping of objects in the Oort cloud was 20%, so this implies that many tens of Earth masses had to have been ejected from the primordial Kuiper belt, consistent within uncertainties with the 35 $M_E$ assumed by the Nice model (Tsiganis et al., 2005).

We assume the primordial Kuiper belt held 29 $M_E$ in 'small' KBOs up to hundreds of km in diameter. We estimate it had ~$6.7 \times 10^{11}$ objects with $D$ > 1 km and $1.3 \times 10^{13}$ objects with $D$ > 50 m. We assume that about 20% of them, or 5.8 $M_E$, were trapped in the Oort cloud and that today it contains $1.3 \times 10^{11}$ objects with $D$ > 1 km and $2.6 \times 10^{12}$ objects with $D$ > 50 m. We likewise estimate that 80%, or $5.3 \times 10^{11}$ objects with $D$ > 1 km, and $1.0 \times 10^{13}$ objects with $D$ > 50 m, would have been ejected to interstellar space, We assume similar ejection efficiencies applied to collisional fragments.

## 2.4   $H_2O$ and $N_2$ ice surface fragments escaping the Solar System

In order for ice fragments to populate interstellar space, they must not be sublimated due to irradiation by Sunlight, nor collide with other objects, before they are scattered out of the Solar System by Jupiter. Fragments are assumed to be generated as Neptune migrates and dynamically excites KBOs so that they collide; as such, the fragments are likely to be part of the scattered disk population once they are generated. We assume they have typical semi-major axes of ~$10^2$ au, with eccentricities that evolve over time so that their perihelia are lowered and they encounter Jupiter.

Survival against sublimation is not ensured because Jupiter resides at ~5 AU and so fragments must approach at least this close to the Sun to be ejected. According to calculations by Jackson & Desch (2021), passage of an $N_2$ ice fragment from great distances to 5 AU and back will cause the diameter to decrease by about 1.4 m. Accounting for the lower luminosity of the Sun in the first ~$10^8$ yr of the Solar System, we estimate each perihelion passage at 5 AU will decrease an $N_2$ ice fragment's diameter by 1.0 m. Fragments probably do not need to undergo many perihelion passages before they encounter Jupiter, though. According to calculations by Wyatt et al. (2017), objects orbiting at 5 AU would be ejected by Jupiter within ~$10^5$ yr; objects with semi-major axes > 100 AU have 1/20 the orbital energy and potentially could be ejected within ~5000 yr. As the orbital period is ~1000 yr, objects may see only 5 perihelion passages, and be eroded only ~5 m in diameter before being ejected. The number of objects (per size bin) that start at diameters 55 m and shrink by 5 m to 55 m is probably only a factor $(55/50)^{-5}$ ~0.6 × the number that started at 50 m, so the number of $N_2$ ice fragments ejected from the Solar System would perhaps be decreased by about a factor of ~1.6 due to thermal processing.

A second requirement is that the fragments do not collide with other KBOs, which would potentially destroy them; or even with each other, which may maintain their compositional distinctiveness but shatter them into smaller, more easily destroyed fragments. Objects are ejected on relatively short timescales, possibly $10^4$ yr (see above), no more than $10^5$ yr (Wyatt et al., 2017), once they are on Jupiter-crossing orbits, but it should take the *e*-folding timescale (~50 Myr) for the eccentricities of objects to evolve onto Jupiter-crossing orbits. Following the discussion of §2.1, the probability of colliding in 50 Myr with the 4.3 $M_E$ of Plutos with radii 1200 km and densities 1.8 g cm$^{-3}$ is $9.0 \times 10^{-4}$. Based on our assumed size distribution of KBOs, we calculate the cross-sectional area-to-mass ratio is equivalent to a diameter of 23 km. If small KBOs have a total mass ~29 $M_E$, diameters 23 km (radii 12 km) and densities 1 g cm$^{-3}$, then the probability of colliding with these objects over 50 Myr is 1 - exp[-($9.0 \times 10^{-4}$) (1.8/1.0) (1200/12) (29/4.3)] = 66%. Likewise, the size distribution of small fragments, with d$N$/d$D \sim D^{-6}$, has an area-to-mass ratio equivalent to objects with diameter 1.5 $D_{min}$, or about 75 m. If fragments have a total mass ~0.074 $M_E$, diameters 75 m, and densities 1 g cm$^{-3}$, then the probability of one fragment colliding with another over 10 Myr (if fragments build up linearly in time, half of all collisions between fragments occur after 40 Myr, at which point the density is 0.8× the final density) is $1 - \exp[-(9.0 \times 10^{-4})$ (1.8/1.0) (1200/0.0375) (0.074×0.8/4.3)(10/50)] = 13%. We therefore calculate that about 61% of fragments will collide with each other or with other KBOs over 50 Myr, and 29% will be ejected.

Combining these probabilities, we estimate that about 18% of $N_2$ ice fragments (and 29% of $H_2O$ ice fragments) could survive against collisions and evaporation long enough to be ejected by Jupiter. The overall mass of fragments would be reduced by an order of magnitude, and the fraction of fragments that are $N_2$ ice fragments would be reduced slightly, but would remain about half. We assume these reductions going forward. Specifically, we assume 0.017 $M_E$ of ice fragments, with 1/2 (about 0.008 $M_E$) of them being $N_2$ ice fragments. Assuming a typical diameter ~70 m, we estimate $\sim 5.6 \times 10^{14}$ small collisional fragments, half of them $N_2$ ice.

Alternatively, passage by a nearby star may strip objects scattered by Neptune into the Oort cloud. These objects would not have entered the inner Solar System and would not have suffered evaporation. It is not clear what fraction of the objects initially in the Kuiper belt could escape through this channel, but it might have been a larger fraction than the ~30% escape probability we estimate. While we do not favor this channel for ejection of most objects from the Solar System, it may be more common in other stellar systems.

2.5  $H_2O$ and $N_2$ ice surface fragments and comets/KBOs in the Oort cloud

A key output of our model is the ratio between $N_2$ ice fragments and regular comets among the ejected bodies, which should match the ratio in the Sun's Oort cloud. Here we assess the probability that $N_2$ ice fragments may reside in the Oort cloud and be observed as long-period comets. We note that the Oort cloud is probably populated by interactions with Neptune, so loss of fragments by collisions may be important, but not by thermal processing.

We calculate that comets and KBOs vastly outweighed the $H_2O$ and $N_2$ ice fragments in the early Solar System by mass, by a factor of (29 $M_E$/0.074 $M_E$) ~400. Comets and KBOs with $D > 1$ km also vastly outnumbered the ice fragments with $D > 1$ km, by a factor $(6.7 \times 10^{11})$ / $(8 \times 10^8)$ ~80. However, because of the steep slope of the fragment size distribution and the very shallow slope of the comet size distribution, among all objects with $D > 50$ m, collisional fragments may have outnumbered comets in the early Solar System by as much as a factor of $(2.7 \times 10^{15})$ / $(1.3 \times 10^{13}) \sim 200$.

If such objects could survive in the Oort cloud, they could be seen today as unusual long-period comets, 1/2 of them likely being $H_2O$ ice-rich fragments of KBO surfaces, and 1/2 being $N_2$ ice fragments. Survival of fragments is not guaranteed, though, as comets in the Oort cloud, beyond the heliosphere boundary at ~100 AU, would be eroded by GCRs. Over the 4.5 Gyr lifetime of the Solar System, $N_2$ ice fragments would have eroded as much as 260 m in radius, and $H_2O$ ice fragments as much as 30 m (Jackson & Desch, 2021), implying that objects with initial $D < 60$ m (if $H_2O$) or $D < 520$ m (if $N_2$) could not survive to the present day to be observed as long-period comets, although fragments with initial $D > 1$ km could. This will lead to an uncertain reduction in the number of fragments among long-period comets, but if the fragment size distribution has a slope of $q=6$, then fragments with $D > 1$ km would be only ~$20^{-5} = 3.1 \times 10^{-7}$ times as numerous as those with $D > 50$ m, and collisionally generated fragments would therefore represent at most $(0.2 \times 2.7 \times 10^{15} \times 3.1 \times 10^{-7}) / (0.2 \times 6.7 \times 10^{11}) \sim 0.13\%$ of all long-period comets > 1 km in diameter, 1/2 of them, or 0.07% (one out of 1400) being $N_2$ ice. This ~1 km size range represents the peak contribution of $N_2$ ice fragments in the present day Oort cloud, since the steep fragment population falls off rapidly at larger sizes, while smaller fragments will not have survived for 4.5 Gyr.

All in all, we estimate that the Sun's Oort cloud holds ~$1.3 \times 10^{11}$ comets / small KBOs with $D > 1$ km (and $2.6 \times 10^{12}$ with $D > 50$ m), having typical size ~few km (Boe et al., 2019), comprising 5.8 $M_E$ (plus about 1600 larger KBOs the sizes of Pluto and Gonggong, with mass 1.2 $M_E$). It originally held ~$2.0 \times 10^{14}$ $N_2$ ice fragments with $D > 50$ m, with of typical size ~70 m, comprising 0.006 $M_E$; but over time, the number of $N_2$ ice fragments was reduced (by GCR erosion) to just those with $D > 1$ km, numbering only about ~$1.7 \times 10^8$. We estimate ~0.07% of all long-period comets with $D > 1$ km may be $N_2$ ice fragments of KBO surfaces. We revisit this point below (§3.3).

## 2.6    $H_2O$ and $N_2$ ice surface fragments and comets/KBOs in interstellar space

Under the assumption that what the Solar System ejected is typical for the numbers and types of objects ejected from other stellar systems (per 1 $M_\odot$ of star), we calculate the frequencies of different objects in the interstellar medium.

At the time of the dynamical instability, we estimate the primordial Kuiper belt contained ~$1.3 \times 10^{13}$ comets/small KBOs with $D > 50$ m, comprising 29 $M_E$, and ~$2.7 \times 10^{15}$ small collisional fragments (1/2 of them $N_2$ ice), comprising 0.074 $M_E$. Only ~29% of fragments survived evaporation or collisions before being ejected. Based on the efficiencies above, we estimate 80% of each population—about $4.9 \times 10^{14}$ fragments and $1.0 \times 10^{13}$ comets/small KBOs—were ejected to interstellar space. If this is the typical number ejected per 1 $M_\odot$ of star, the number density of ice fragments in the ISM would be $5 \times 10^{13}$ pc$^{-3}$, . These would have typical diameters ~70 m (with half of those being $N_2$ ice and half being $H_2O$ ice). This result is uncertain by factors of a few, depending on assumptions about the fragment size distribution, and especially the minimum diameter of fragments, but compares very favorably to the number densities inferred by Poretegies Zwart et al. (2018) and Widmark and Monari (2019). They estimated a number density in the range $3.5 \times 10^{13}$ pc$^{-3}$ to $2 \times 10^{15}$ pc$^{-3}$, favoring $7 \times 10^{14}$ pc$^{-3}$.

If what was ejected from the Solar System is typical, we estimate that collisional fragments would have dominated the population of small interstellar objects with $D > 50$ m, as they would outnumber comets/small KBOs ($D > 50$ m) by a factor of ~50, and larger comets/KBOs ($D > 1$ km) by a factor of ~900, at least at first, while larger fragments ($D > 1$ km) would be much rarer and less common than comets of comparable size. Roughly 1/2 of these fragments from differentiated KBOs would be $N_2$ ice, and 1/2 $H_2O$ ice. This ratio would decrease, as $H_2O$ ice fragments and especially $N_2$ ice fragments would be destroyed by GCRs. Based on our the GCR erosion rate calculated by Jackson & Desch (2021), $N_2$ ice fragments with $D = 50$ m older than 0.9 Gyr would have been destroyed. Likewise, $H_2O$ ice fragments, and comets/KBOs, with $D = 50$ m older than 3.1 Gyr, would have been destroyed. Using the star formation history of Mor et al. (2019), only 5% of stars have formed in the last 0.9 Gyr. and only 35% of stars have formed in the last 3.1 Gyr, suggesting that the fraction of $N_2$ ice fragments relative to $H_2O$ ice fragments, among objects in the ISM, is a factor of 7 lower than the ratio in material immediately after ejection. Among all surviving small bodies ($D > 50$ m), only ~5% should be $N_2$ ice fragments. The ratio of fragments to larger ($D > 1$ km) comets and KBOs, which would survive to the present day after having been ejected throughout Galactic history, would be reduced by a factor ~4. Fragments with $D > 50$ m would outnumber comets with $D > 1$ km by a factor of only 200. After accounting for GCR erosion while in the ISM, the final density of $N_2$ ice fragments would be $1.4 \times 10^{12}$ pc$^{-3}$, of $H_2O$ ice fragments would be $1.1 \times 10^{13}$ pc$^{-3}$, of small KBOs/comets $> 50$ m would be $1.2 \times 10^{12}$ pc$^{-3}$, and of KBOs/comets $> 1$ km would be $6.4 \times 10^{10}$ pc$^{-3}$. The number density of all interstellar objects would be ~ $1.4 \times 10^{13}$ pc$^{-3}$. While slightly below the lower limit of $3.5 \times 10^{13}$ pc$^{-3}$ considered statistically likely by Portegies Zwart et al. (2018) and Widmark and Monari (2019), we consider the match close enough that collisional fragments could plausibly explain interstellar objects like 'Oumuamua. Among objects $> 50$ m in the ISM, about 10% could be expected to be $N_2$ ice fragments. The average age of the $N_2$ ice fragments would be ~0.5 Gyr, the average age of the $H_2O$ ice fragments would be ~2 Gyr, and the average age of other objects would be ~5 Gyr. The average sizes of the collisional fragments would be $D$ ~70 m at first, but significantly eroded by passage throught the ISM over many Gyr. The average sizes of the comets/KBOs would be closer to $D$ ~3 km, as is typical for comets. Fragments would make up ~90% of all interstellar objects with $D > 50$ m, but a very small fraction (~0.1%) of objects with $D > 1$ km.

We note that our hypothesized size distribution involving multiple populations—ejection of about 23.2 $M_E$ of KBOs ranging from < 1 km to hundreds of km in radius, plus 4.8 $M_E$ of Gonggong-like and Pluto-like dwarf planets ~$10^3$ km in radius, and ~0.015 $M_E$ (~0.1% of the ejected mass) of 0.05-1 km $N_2$ and $H_2O$ ice fragments generated by the impacts of the first objects onto the second—is not dissimilar to mixes of populations considered by 'Oumuamua ISSI Team et al. (2019). For example, their scenario "*b2*" entails a KBO population with a primordial size distribution extending from $D$=100 m up to 100 km, plus 3% of the mass in $D >$ 50 m fragments, the two populations comprising up to 3 $M_E$ pc$^{-3}$, or ~25 $M_E$ per 1 $M_\odot$ of star. The various distributions investigated by 'Oumuamua ISSI Team et al. (2019) if scaled to 29 $M_E$, imply numbers (per 1 $M_\odot$ of star) $7 \times 10^{12}$ ($D_{min}$=200 m, d$N(>M)$/d$M = M^{-0.6}$), or $2 \times 10^{15}$ ($D_{min}$=100 m, d$N(>M)$/d$M = M^{-0.6}$), or $6 \times 10^{16}$ (extrapolating from the boulder size distribution on comet 67P/Churumyukov-Gerasimov).

## 2.7 Summary

We conclude that if the Solar System was typical in the types and numbers of objects it ejected, then the number density of collisional fragments in the ISM, $\sim 1.4 \times 10^{13}$ pc$^{-3}$, is reasonably in line with the numbers of objects inferred by Portegies Zwart et al. (2018) and Widmark and Monari (2019). It would be somewhat unusual ($\sim 10\%$) for such a fragment to be $N_2$ ice, as we infer 'Oumuamua to be, but nevertheless, collisionally generated ice fragments from planetary surfaces are, we predict, an abundant type of interstellar object, and almost an order of magnitude more abundant than traditional comets/small KBOs.

In **Table 1** we list our predictions for how much material was ejected from the Solar System into the Oort cloud, and into interstellar space. We also list our predicted nmber densities of comets and $H_2O$ and $N_2$ ice fragments, taking into account that GCR erosion means only $\sim 5\%$ of $N_2$ fragments and $\sim 35\%$ of $H_2O$ ice fragments formed over Galactic history would survive to the present day, and extrapolating to a density in the ISM by multiplying the number of objects per 1 $M_\odot$ by an average stellar density 0.12 $M_\odot$ pc$^{-3}$.

**Table 1**: Number and masses of ordinary comets/KBOs, and $H_2O$ and $N_2$ ice fragments of various sizes. We assume 29 $M_E$ with the size distribution of the dynamically cold population. Only 29% of $H_2O$ and 18% of $N_2$ ice fragments assumed to survive long enough to be ejected. Of those ejected, 80% escape, 20% are emplaced in Oort cloud. Bodies > 1 km survive against GCRs, but those $\sim 50$ m do not. In the ISM, among bodies > 50 m, only the 35% of $H_2O$ and 5% of $N_2$ bodies were ejected recently enough to exist. We assume 0.12 $M_\odot$ pc$^{-3}$ in the ISM.

| | Large KBOs | Comets / small KBOs | | $H_2O$ ice fragments | | $N_2$ ice fragments | |
|---|---|---|---|---|---|---|---|
| | | $D > 50$ m | $D > 1$ km | $D > 50$ m | $D > 1$ km | $D > 50$ m | $D > 1$ km |
| In Kuiper belt after Neptune's migration | 8000 | $1.3 \times 10^{13}$ | $6.7 \times 10^{11}$ | $1.1 \times 10^{15}$ | $3.5 \times 10^{8}$ | $1.6 \times 10^{15}$ | $5.0 \times 10^{8}$ |
| Put in Sun's Oort cloud 4.5 Gyr ago | 1600 | $2.6 \times 10^{12}$ | $1.3 \times 10^{11}$ | $6.5 \times 10^{13}$ | $7.0 \times 10^{7}$ | $5.7 \times 10^{13}$ | $1.0 \times 10^{8}$ |
| In Oort Cloud today | 1600 | - | $1.3 \times 10^{11}$ | - | $7.0 \times 10^{7}$ | - | $1.0 \times 10^{8}$ |
| Ejected from Sun 4.5 Gyr ago | 6400 | $1.0 \times 10^{13}$ | $5.3 \times 10^{11}$ | $2.6 \times 10^{14}$ | $2.8 \times 10^{8}$ | $2.3 \times 10^{14}$ | $3.9 \times 10^{8}$ |
| In ISM today (per pc$^3$) | 770 | $1.2 \times 10^{12}$ | $6.4 \times 10^{10}$ | $1.1 \times 10^{13}$ | $3.4 \times 10^{7}$ | $1.4 \times 10^{12}$ | $4.7 \times 10^{7}$ |

## 3 Discussion

### 3.1 Comparison between 1I/'Oumuamua and 2I/Borisov

The model presented above is consistent with the differences between 1I/'Oumuamua and 2I/Borisov, starting with their velocities relative to the solar system. Per the above discussion (§2.5), the average age of $N_2$ ice fragments should be ~0.5 Gyr, whereas $H_2O$ ice-dominated comets like 2I/Borisov should have mean ages ~5 Gyr. Jackson & Desch (2021) estimate 'Oumuamua itself is likely ≈ 0.4-0.5 Gyr old, consistent with the average age of an $N_2$ ice fragment. 'Oumuamua's velocity with respect to the LSR is estimated to be between about 3 and 11.5 km/s (Feng and Jones, 2018; Hallatt and Wiegert, 2020). These velocities are typical of young stars < 2 Gyr old; after travelling through the Galaxy for longer than this, stars typically scatter off each other and acquire much higher random velocities, such as the ~20 km s$^{-1}$ peculiar velocity of the Sun (Almeida-Fernandes and Rocha-Pinto, 2018). More precisely, Almeida-Fernandes and Rocha-Pinto (2018) placed an upper limit to its age of $1.9 - 2.1$ Gyr, depending on the kick it received upon ejection. This implies both that 'Oumuamua derived from a system < 2 Gyr old (the dynamical instability in the Solar System probably took place at ~0.1 Gyr after formation), and that it has been traveling through the Galaxy for < 2 Gyr (we favor ~0.4-0.5 Gyr). Our derived age of 0.4 Gyr is long enough for 'Oumuamua to have been significantly eroded during passage in the interstellar medium, but not long enough to acquire a significant random velocity. 2I/Borisov, in contrast, if it is billions of years old, should have a velocity difference with respect to the LSR of tens of km/s (Feng and Jones, 2018). This is consistent with the observed 35 km/s (Feng and Jones, 2018; Hallatt and Wiegert, 2020).

At the time of ejection from the Solar System, fragments made up only 0.015 $M_E$ of ejected material, and only 1/2 of that was $N_2$ ice, whereas comets/small KBOs made up 23.2 $M_E$ of material. It is worth asking how statistically likely or unlikely it is that a ~25 m-diameter $N_2$ ice fragment and a ~500 m-diameter, apparently normal, comet should represent the first two confirmed interstellar objects.

In surveys sensitive to objects the size of 'Oumuamua, collisional fragments should be the most common object. We calculate their combined number density in the ISM to be ~$1.2 \times 10^{13}$ pc$^{-3}$, compared to $1.2 \times 10^{12}$ pc$^{-3}$ for comets. Of these fragments, about 90% are $H_2O$ ice fragments (~$1.1 \times 10^{13}$ pc$^{-3}$) and 10% are $N_2$ ice fragments (~$1.4 \times 10^{12}$ pc$^{-3}$), due to $N_2$ ice being more easily eroded by GCRs in the ISM. On this basis we might conclude that a fragment is likely to be observed, but there would be only a 10% probability that we should see an $N_2$ ice fragment. However, it is worth noting that 'Oumuamua only brightened to magnitude +19.7 and was at the limits of detection; it may have been observed only because of its high albedo, which we infer was ~0.64, like the surface of Pluto. If it were more like the $H_2O$ ice-covered surface of Charon, with albedo about ~0.25 (Buratti et al., 2017), it would have had to be 1.6× larger in diameter (~40 m) to be detected, and if the size frequency distribution of the fragments has slope $q \approx 6$, such objects would be rarer by a factor of ~0.10. That is, the number density of observable $N_2$ ice fragments would be the same, ~$1.4 \times 10^{12}$ pc$^{-3}$, but the number density of observable $H_2O$ ice fragments would effectively be ~$1.1 \times 10^{12}$ pc$^{-3}$. This increases the probability that the first small fragment to be detected would be $N_2$ ice rather than $H_2O$ ice, to about 50%.

We calculate that the number density in the ISM of large ($D > 1$ km) comets like 2I/Borisov should be ~$6.5 \times 10^{10}$ pc$^{-3}$, about 0.5% of the number density of all small ($D > 50$ m) objects, mostly collisional fragments, suggesting that objects like 2I/Borisov are much rarer than collisional fragments. However, detection of 'Oumuamua occurred only because it passed within ~0.2 AU of Earth. In contrast, 2I/Borisov was detected when 3.7 AU from Earth, after reaching magnitude +15, in part because it had a large coma. Borisov is estimated to have a diameter 0.5 km (Jewitt et al., 2020). Lacking dust, collisional fragments of differentiated KBOs are not as detectable as comets; but we estimate that Borisov would have been detected from Earth once it approached within 3 AU of the Sun and sported a visible coma. This increases the probability that a comet like 2I/Borisov would be detected, perhaps by a factor ~$10^3$, making it comparably probable to be seen, despite being a rarer object than collisional fragments. Portegies Zwart et al. (2018) explicitly state that prior estimates of interstellar objects (e.g., Engelhardt et al., 2017) were predicated on the expectation that such objects would sport comet-like comas and be easier to detect.

It is difficult to assess the relative frequencies of such objects from observations, as they involve different techniques with different selection biases, but both comets and KBO surface collisional fragments would seem likely to be observed in a 5-year span of time, with collisional fragments dominating by number, but comets being much more easily detected by virtue of developing a coma of dust and gas. Among collisional fragments, most would seem to be $H_2O$ ice, and the fraction that are $N_2$ ice starts at ~1/2, but drops to 10% as $N_2$ ice fragments are destroyed; but the higher albedo makes the smallest-sized $N_2$ ice fragments easier to detect, and perhaps 50% of observed fragments will be $N_2$ ice.

We conclude that the discovery of objects like 1I/'Oumuamua and 2I/Borisov are not inconsistent with their expected rates of occurrence.

## 3.2 Could other $N_2$ ice fragments remain in our Solar System?

We have argued that the number of $N_2$ ice fragments with $D > 1$ km, capable of surviving against GCR erosion to the present day, put into the Oort cloud was about $1 \times 10^8$, compared to about $1.3 \times 10^{11}$ comets/small KBOs with $D > 1$ km. About 0.08%, or one in 1300, of long-period comets could be $N_2$ ice fragments. Among fragments with $D > 0.5$ km, potentially survivable against GCR erosion for 4.5 Gyr, the proportion is 1.5%, or one in 70. Given that thousands of comets have been observed, it is intriguingly possible that ~0.1-1%, or at least one, long-period comet may be a collisional fragment.

The recent long-period comet C/2016 R2 may represent one such object. This comet was discovered in January 2018 and its composition was measured by millimeter spectroscopy when it was at 2.8 au (Biver et al., 2018). Based on its unusual chemistry, Biver et al. (2018) suggested it may be a collisional fragment of a KBO. Unlike other comets, OH (derived from $H_2O$) was not detected, and low upper limits on S species were placed. There was very little dust production. The most abundant species were CO and $N_2$, with an abundance ratio of 100:8. The outgassing rate ~$4 \times 10^{28}$ CO molecules s$^{-1}$ suggests a body 3 km in diameter (Biver et al., 2018). The dominance of $N_2$ among the observed nitrogen-bearing species is unusual. The $N_2$/HCN ratio in C/2016 R2 is $10^4$ × higher than other comets. There are no hints of other N-bearing species (e.g., $NH_2$), and it is inferred that all of the N is in the form of $N_2$ ice. While this body has a much higher proportion of CO to $N_2$ than the pure $N_2$ ice composition we infer for 'Oumuamua, the $N_2$/CO ratio in most comets is usually much smaller (e.g., 6 x $10^{-5}$ in comet Hale-Bopp: Cochran et al., 2000); few × $10^{-3}$ in 67P/Churyumov-Gerasimenko: Rubin et al., 2015) than the ~0.08 observed in C/2016 R2. While 'Oumuamua did not contain such large amounts of CO, this is a compound that is found on KBO surfaces.

According to Biver et al. (2018), the only other comets exhibiting such large $N_2$ and CO production and high $N_2$/CO ratios and low dust production were C/1908 R1 Morehouse and C/1961 R1 Humason; the comets 29P/Schwassmann-Wachmann 1 and C/2002 $VQ_{94}$ LINEAR were similar chemically, but dust-rich. We therefore infer that 1, possibly 3 out of all the known comets may in fact be $N_2$-rich collisional fragments of the icy crust of KBOs. The occurrence rate of such chemically distinct comets would thus seem to be consistent with ~0.1%.

We believe the suggestion of Biver et al. (2018), that C/2016 R2 is a collisional fragment from a KBO surface, has merit and is worth further investigation to compare it and similar comets to 'Oumuamua. An assessment of their frequency among long-period comets could provide constraints on the number of such objects ejected from the Solar System into interstellar space.

### 3.3 Universality of collisional fragment ejection among stellar systems

The frequency of detection of objects like 'Oumuamua is matched by our predictions, but *only* if the types and total masses of materials ejected by the Sun are typical (per stellar mass) for stars overall. 'Oumuamua would not be a likely event if most stars didn't also undergo a Nice model-like dynamical instability. The Solar System ejected perhaps 28 $M_E$ of material during its dynamical instability that depleted the primordial Kuiper belt. The solid mass incorporated into the Solar System planets exceeded about 80 $M_E$ and the solid mass in the Sun's protoplanetary disk probably approached 200 $M_E$ (Desch, 2007), so ejection of 28 $M_E$ would represent a loss of maybe 15% of the planetary mass by dynamical instability. Since it is not possible to eject a much larger fraction of the planetary mass in any system, it must be the case that the majority of systems eject a comparable amount of material.

The mass of planets and planet-forming materials in a stellar system probably scales with the stellar mass. Protoplanetary disk masses tend to scale with stellar masses (Andrews et al., 2013), and apparently so do the masses of planets that form. For example, the TRAPPIST-1 system around an 0.08 $M_\odot$ star contains at least 5 $M_E$ of rocky or icy material in planets (Grimm et al., 2018), or ~60 $M_E$/$M_\odot$, compared to ~80 $M_E$ / $M_\odot$ in the Solar System (Desch, 2007). If similar numbers of $N_2$ ice fragments are ejected per mass of star, then we predict that the number density of small bodies (> 50 m) in the ISM should be ~1 × $10^{13}$ $pc^{-3}$, a fraction ~10% of these being $N_2$ ice collisional fragments. If a small fraction of stars were to undergo a dynamical instability in their Kuiper belts, or if $N_2$ ice wasn't produced on the surfaces of KBOs in most stars, then the density of 'Oumuamua-like objects would be much lower than we predict. Conversely, the number could be higher if ejection were more efficient in other systems.

Dynamical instabilities have long been thought to be common among exoplanetary systems (Rasio and Ford, 1996; Weidenschilling and Marzari, 1996; Lin and Ida, 1997; Raymond et al., 2010, 2011, 2012), and there are reasons to assume they would be universal. The state of lowest energy in a disk of planetoids is one in which some fraction of the mass (e.g., Neptune) migrates to large semi-major axis, carrying the system angular momentum with it, while other planets (e.g., Jupiter, Saturn and Uranus) migrate inward, as in the Nice model (Tsiganis et al., 2005). The exchange of orbital energy and angular momentum is mediated through smaller planetesimals (e.g., KBOs), a large fraction of which can be ejected (Wyatt et al., 2017). As long as multiple planets exist, dynamical instabilities depleting a Kuiper belt-like reservoir can be expected to occur, on timescales comparable to the age of the system itself (Laskar 1996; Volk and Gladman, 2015).

Whether the planetesimals scattered by the dynamical instability can escape depends on whether a stellar system hosts a planet capable of ejecting fragments to interstellar space. If a Jupiter-mass planet were required, this would cut down the efficiency of $N_2$ ice fragment ejection, perhaps by an order of magnitude, as the occurrence rate of Jupiter analogs in FGK systems is estimated as only $\approx 6.7\%$ (Wittenmyer et al., 2020), possibly up to $\approx 15\%$ (Gaudi et al., 2008). However, both simple considerations of the Safronov number and more detailed calculations (for example, Wyatt et al. (2017) show that ejection generally requires only a Neptune-mass planet around a G star with mass $\approx 1~M_\odot$, a Saturn-mass planet around an A star with mass $\approx 2~M_\odot$, or a planet with a few Earth masses around an M star with mass $\approx 0.08~M_\odot$. Given the relative abundance of Neptune-sized planets around FGK stars (Batalha, 2014), and of super-Earths around M stars, the efficiency of ejection may not be significantly curtailed around these stars. It is unclear whether A stars generally host Saturn-sized planets, and the commonality of debris disks around A stars suggests planet formation might not be as efficient around these stars (Lisse et al., 2019). Integrating the initial mass function of Kroupa (2001), about 34% of the mass of all stars is in M stars with mass $< 0.8~M_\odot$, about 50% in stars of type A or earlier, with masses $> 2~M_\odot$, and only about 16% of the mass is in FGK stars. It would seem more probable that objects like 'Oumuamua could be ejected from FGK and M star systems than from A star systems. However, even if fragments could only be ejected from stars of spectral type G and later, this would only cut down the frequency of collisional fragments by a factor of 2. It is worth noting that a number of physical processes have been identified that could lead to fragments being ejected from a young stellar system (Ćuk, 2018; Wyatt et al., 2017; Raymond et al. 2018a,b; 'Oumuamua ISSI Team et al., 2019), including gravitational interactions with stars in a stellar birth cluster, or interaction with a binary star (Holman and Wiegert, 1999). These factors would increase the rate of ejection.

A further investigation of the efficiency of ejection of $N_2$ ice fragments from stellar systems with host stars of different mass is beyond the scope of the current paper. The frequency of objects like 'Oumuamua is explained if other stellar systems eject objects with the same efficiency per stellar mass as the Sun, and indeed this process must be fairly universal in order for detection of an $N_2$ ice fragment to be a probable event; but there are no *a priori* reasons why this assumption would not hold to within a factor of a few, and remain within the observational constraints.

The number density of $N_2$ ice fragments interstellar objects we predict, extrapolating from the Sun to other stars, is $\sim 1.4 \times 10^{13}$ pc$^{-3}$. This is slightly lower than the range of values considered plausible by Portegies Zwart et al. (2018), $3.5 \times 10^{13}$ pc$^{-3}$ to $2 \times 10^{15}$ pc$^{-3}$, which is the 95% confidence level based on a single detection in a volume 0.08 au$^3$ and the assumption that the object could be detected only if it had 'Oumuamua's magnitude, +20. We note that the limiting magnitude of the survey is closer to +22, which suggests a survey volume larger by a factor of 6, in which case the (95% confidence level) lower limit to the number density would be ~6 $\times 10^{12}$ pc$^{-3}$. Besides the uncertainty in determining a number density from the single detection of an object like 'Oumuamua, there are considerable uncertainties in extrapolating from the Sun to other stellar systems. Given these order-of-magnitude uncertainties, we consider the match to be reasonable, and that it is not implausible that 'Oumuamua could be an $N_2$ ice fragment arising from collisions with differentiated KBOs.

## 4    Summary

We have demonstrated that 'Oumuamua is consistent with a collisional fragment of $N_2$ ice from a differentiated KBO, and that the frequency of such bodies is consistent with observations, provided most stellar systems create and eject such fragments with the efficiency with which our solar system did, > 4 Gyr ago. This strongly implies that the relevant processes are somewhat universal among stellar systems. This includes: the rapid (< 10-100 Myr) formation and differentiation of thousands of differentiated planets like KBOs, that emplace $N_2$ ice on their surfaces as Pluto and Triton have; a relatively late (> 10-100 Myr) dynamical instability that would erode their surfaces and create ~0.1 $M_E$ of collisional fragments, including a high fraction of $N_2$ ice fragments; and the ability to eject most of these fragments from the system, by giant planets, binary companion, or passing star (Wyatt et al., 2017). Without all of these processes occurring simultaneously, the occurrence of an $N_2$ ice fragment like 'Oumuamua would be improbable. This strongly constrains models of planet formation, including models considering a variety of stellar masses.

Our scenario predicts that interstellar objects should be dominated by small ($D > 50$ m) collisional fragments, with about 1/2 of them being $N_2$ ice fragments, the other 1/2 being presumably $H_2O$ ice fragments, at the time of ejection. Both would be eroded by GCRs during transit through the ISM, at rates of tens of meters per Gyr. We calculate that $N_2$ ice fragments would typically survive ~1 Gyr, and $H_2O$ ice fragments typically ~3 Gyr, and their mean ages would be ~0.5 Gyr and ~2 Gyr. Because of this differential survival rate, $N_2$ ice fragments may make up only ~10% of collisional fragments reaching our Solar System. However, because of their high albedo, it is possible to detect smaller $N_2$ ice fragments, so that there is a closer to ~50% probability that the first interstellar collisional fragment observed would be $N_2$ ice like 1I/'Oumuamua. The number of collisional fragments exceeds the number of regular comets/small KBOs like 2I/Borisov in the interstellar medium, but the fact that the latter objects develop comas and brighten means observations of them may outnumber observations of collisional fragments. The statistics of interstellar objects that we predict can be tested when the *Vera C. Rubin Observatory* (formerly *Large Synoptic Survey Telescope*), which may begin discovering interstellar objects at the rate ~1/year beginning in 2022 (Cook et al., 2016).

Based on the mean ages of $N_2$ ice fragments, we view it as likely that 'Oumuamua has been traveling through interstellar space for a significant fraction of ~1 Gyr. This timeframe is consistent with the maximum transit time ~2 Gyr to avoid acquiring too high a velocity with respect to the LSR (Almeida-Fernandes and Rocha-Pinto, 2018). It may not be possible to constrain the transit time more precisely, but the transit time suggested by Jackson & Desch (2021), ~0.4-0.5 Gyr, would be probable, and would allow 'Oumuamua to be ejected from its stellar system with axis ratios 1.7:1 that might be more typical of collisional fragments than the 2.1:1 ratio it is inferred to have upon entry to the Solar System, as well as being consistent with the average age of an $N_2$ ice fragment. An object like 'Oumuamua must be ejected from a young system (< 2 Gyr) to avoid acquiring a large velocity, and probably a very young system $< 10^8$ yr, if the timing of the dynamical instability that depleted the primordial Kuiper belt is typical. Such a young system is likely to be located in a spiral arm of the Galaxy, especially if the instability happens at early times $\sim 10^7$ yr. A transit time $\sim$0.4-0.5 Gyr at 9 km s$^{-1}$ would imply travel over 3.6 kpc, and this distance and the direction of 'Oumuamua's approach would seem consistent with an origin in the Perseus arm.

This sequence of events is summarized in **Figure 1**, which uses the results from this paper and the companion paper (Jackson and Desch 2021). The axes upon ejection from its parent system are consistent with the expected size of collisional fragments, with mean diameter $\sim$70 m, and the axis ratios thought typical of fragments. The loss of about half its mass in the ISM by erosion by Galactic cosmic rays is consistent with transit for $0.4 - 0.5$ Gyr, about half the time thought typical. Other aspects of its evolution within the Solar System are discussed in the companion paper. The extreme mass loss experienced as it passed the Sun accounts for its small size and its extreme axis ratios.

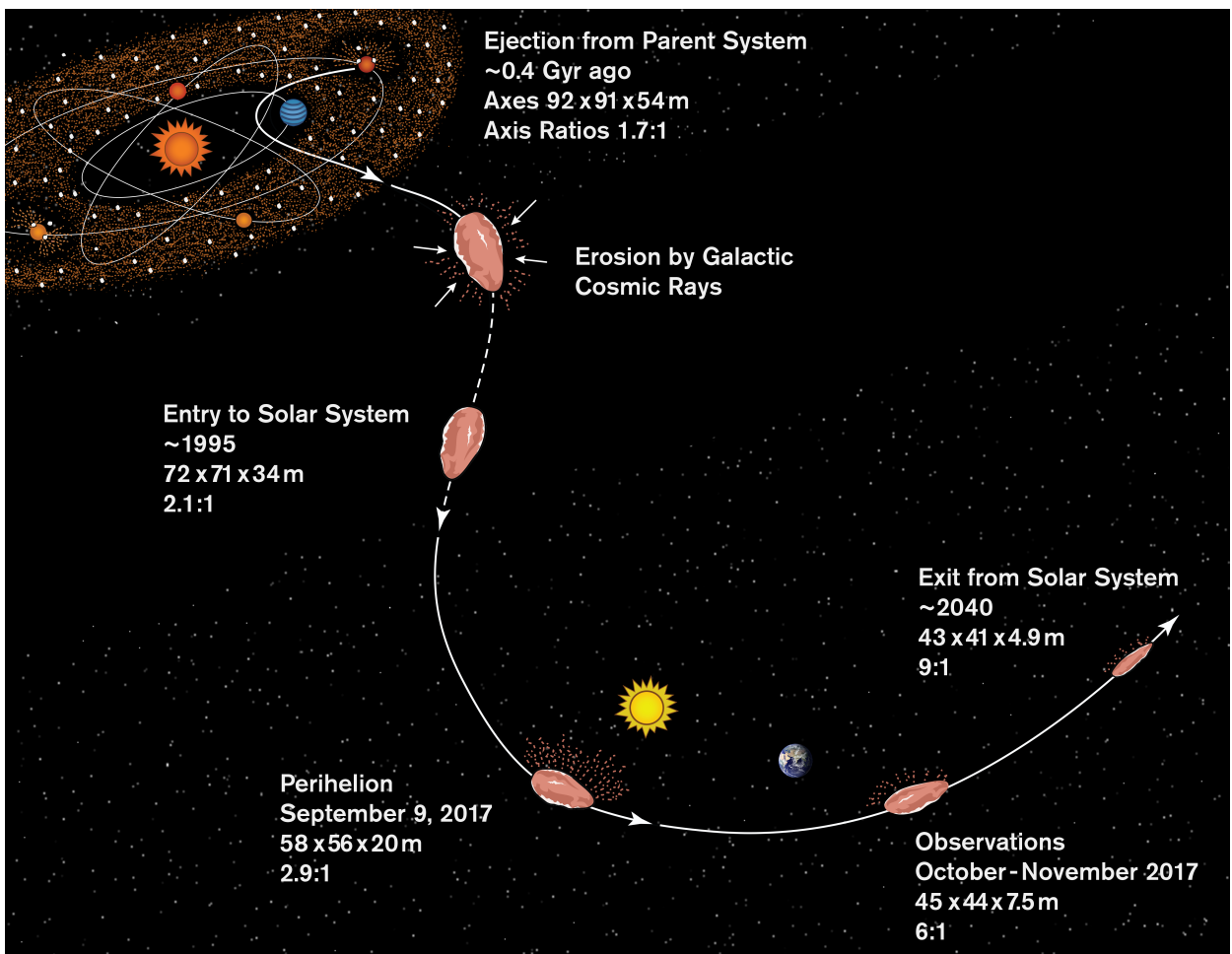

**Figure 1.** A plausible history for 'Oumuamua as a collisional fragment of $N_2$ ice, as hypothesized in this paper and the companion paper by Jackson and Desch (2021). *Credit*: Sue Selkirk.

If 'Oumuamua truly is an $N_2$ ice fragment of a differentiated, Pluto-like object, the greatest implication is that we have observed the composition of a new type of exoplanet, an 'exo-Pluto'. Exoplanet compositions are difficult to observe, information about planetary interiors can only be inferred indirectly from observations of atmospheric gases (e.g., Seager, 2014), or in bulk if material falls onto a white dwarf long after the death of the system (Harrison, 2018; Swan et al., 2019). Observations of objects in their stars' Kuiper belts today may not be possible in the foreseeable future. We predict that collisional fragments of exo-Plutos may pass through the Solar System roughly once per year. The composition of a fragment of an exoplanet brought to within ~0.2 AU of Earth can be determined to much more precise extent than is currently possible for exoplanets around their host stars, using techniques common to cometary observations. Application of these observational techniques from planetary science to interstellar objects may open new vistas on the compositions and formation mechanisms and dynamics of exoplanetary systems.

**Acknowledgments**

We thank Yusuke Fujimoto, Chris Glein, Greg Laughlin, and Darryl Seligman for useful discussions. We thank an anonymous referee and especially the referee Sean Raymond for comments that helped us improve the clarity of the manuscript. The results reported herein benefitted from collaborations and/or information exchange within NASA's Nexus for Exoplanet System Science (NExSS) research coordination network sponsored by NASA's Science Mission Directorate.

Data were not used, nor created, for this research.

# References

Almeida-Fernandes, F., & Rocha-Pinto, H. J. (2018). A kinematical age for the interstellar object 1I/`Oumuamua. *Monthly Notices of the Royal Astronomical Society, 480*(4), 4903-4911. doi:10.1093/mnras/sty2202

Andrews, S. M., Rosenfeld, K. A., Kraus, A. L., & Wilner, D. J. (2013). The Mass Dependence between Protoplanetary Disks and their Stellar Hosts. *The Astrophysical Journal, 771*(2), ID129. doi:10.1088/0004-637X/771/2/129

Bailer-Jones, C. A., Farnocchia, D., Ye, Q., Meech, K. J., & Micheli, M. (2020). A search for the origin of the interstellar comet 2I/Borisov. *Astronomy & Astrophysics, 634*, A14. doi:10.1051/0004-6361/201937231

Barucci, M. A., Dalle Ore, C. M., Perna, D., Cruikshank, D. P., Doressoundiram, A., Alvarez-Candal, A., . . . Nitschelm, C. (2015). (50000) Quaoar: Surface composition variability. *Astronomy & Astrophysics, 584*, A107. doi:10.1051/0004-6361/201526119

Batalha, N. M. (2014). Exploring exoplanet populations with NASA's Kepler Mission. *Proceedings of the National Academy of Sciences, 111*(35), 12647-12654. doi:10.1073/pnas.1304196111

Belton, M. J., Hainaut, O. R., Meech, K. J., Mueller, B. E., Kleyna, J. T., Weaver, H. A., . . . Keane, J. V. (2018). The Excited Spin State of 1I/2017 U1 'Oumuamua. *The Astrophysical Journal Letters, 856*(2), L21. doi:10.3847/2041-8213/aab370

Bialy, S., & Loeb, A. (2018). Could Solar Radiation Pressure Explain 'Oumuamua's Peculiar Acceleration? *The Astrophysical Journal Letters, 868*(1), L1. doi:10.3847/2041-8213/aaeda8

Bierson, C. J., Nimmo, F., Stern, S. A., Olkin, C. B., Weaver, H. A., Young, L., & Ennico, K. (2020). The Plausibility of an Ocean on Pluto Shortly After Accretion. *51st Lunar and Planetary Science Conference* (p. 2326). The Woodlands, Texas: Lunar and Planetary Science Institute.

Biver, N., Bockelée-Morvan, D., Paubert, G., Moreno, R., Crovisier, J., Boissier, J., . . . DiSanti, M. A. (2018). The extraordinary composition of the blue comet C/2016 R2 (PanSTARRS). *Astronomy & Astrophysics, 619*, A127. doi:10.1051/0004-6361/201833449

Boe, B., Jedicke, R., Meech, K. J., Wiegert, P., Weryk, R. J., Chambers, K. C., . . . Waters, C. (2019). The orbit and size-frequency distribution of long period comets observed by Pan-STARRS1. *Icarus, 333*, 252-272. doi:10.1016/j.icarus.2019.05.034

Brasser, R., & Morbidelli, A. (2013). Oort cloud and Scattered Disc formation during a late dynamical instability in the Solar System. *Icarus, 225*(1), 40-49. doi:10.1016/j.icarus.2013.03.012

Brasser, R., Duncan, M. J., & Levison, H. F. (2006). Embedded star clusters and the formation of the Oort Cloud. *Icarus, 184*(1), 59-82. doi:10.1016/j.icarus.2006.04.010

Buratti, B. J., Hicks, M. D., Dalba, P. A., Chu, D., O'Neill, A., Hillier, J. K., . . . Rhoades, H. (2015). Photometry of Pluto 2008-2014: Evidence of Ongoing Seasonal Volatile Transport and Activity. *The Astrophysical Journal Letters, 804*(1), L6. doi:10.1088/2041-8205/804/1/L6

Buratti, B. J., Hofgartner, J. D., Hicks, M. D., Weaver, H. A., Stern, S. A., Momary, T., . . . Olkin, C. B. (2017). Global albedos of Pluto and Charon from LORRI New Horizons observations. *Icarus, 287*, 207-217. doi:10.1016/j.icarus.2016.11.012

Canup, R. M., Kratter, K. M., & Neveu, M. (2020). Formation of the Pluto system. In S. A. Stern, & e. al. (Eds.), *The Pluto system after New Horizons.* in Press.

Chang, H.-K., Liu, C.-Y., & Shang, J.-R. (2016). Upper limits to the number of Oort Cloud objects based on serendipitous occultation events search in X-rays. *Monthly Notices of the Royal Astronomical Society, 462*(2), 1952-1960. doi:10.1093/mnras/stw1781

Clement, M. S., Kaib, N. A., Raymond, S. N., & Walsh, K. J. (2018). Mars' growth stunted by an early giant planet instability. *Icarus, 311*, 340-356. doi:10.1016/j.icarus.2018.04.008

Cochran, A. L., Cochran, W. D., & Barker, E. S. (2000). N+_2 and CO+ in Comets 122P/1995 S1 (deVico) and C/1995 O1 (Hale-Bopp). *Icarus, 146*(2), 583-593. doi:10.1006/icar.2000.6413

Cook, N. V., Ragozzine, D., Granvik, M., & Stephens, D. C. (2016). Realistic Detectability of Close Interstellar Comets. *The Astrophysical Journal, 825*(1), 51. doi:10.3847/0004-637X/825/1/51

Cruikshank, D. P., Roush, T. L., Owen, T. C., Quirico, E., & De Bergh, C. (1995). *The Surface Compositions of Triton, Pluto, and Charon.* Moffett Field, CA: NASA Ames Research Center.

Cruikshank, D. P., Roush, T. L., Owen, T. C., Quirico, E., & de Bergh, C. (1998). The Surface Compositions of Triton, Pluto, and Charon. In B. Schmitt, C. de Bergh, & M. Festou (Eds.), *Solar System Ices* (Vol. 227, p. 655). doi:10.1007/978-94-011-5252-5_27

Ćuk, M. (2018). 1I/'Oumuamua as a Tidal Disruption Fragment from a Binary Star System. *The Astrophysical Journal Letters, 852*(1), L15. doi:10.3847/2041-8213/aaa3db

de Sousa, R. R., Morbidelli, A., Raymond, S. N., Izidoro, A., Gomes, R., & Vieira Neto, E. (2020). Dynamical evidence for an early giant planet instability. *Icarus, 339*, 113605. doi:10.1016/j.icarus.2019.113605

Desch, S. J. (2007). Mass Distribution and Planet Formation in the Solar Nebula. *The Astrophysical Journal, 671*(1), 878-893. doi:10.1086/522825

Desch, S. J., Cook, J. C., Doggett, T. C., & Porter, S. B. (2009). Thermal evolution of Kuiper belt objects, with implications for cryovolcanism. *Icarus, 202*(2), 694-714. doi:10.1016/j.icarus.2009.03.009

Do, A., Tucker, M. A., & Tonry, J. (2018). Interstellar Interlopers: Number Density and Origin of 'Oumuamua-like Objects. *The Astrophysical Journal Letters, 855*(1), L10. doi:10.3847/2041-8213/aaae67

Domokos, G., Sipos, A. Á., Szabó, G. M., & Várkonyi, P. L. (2009). Formation of Sharp Edges and Planar Areas of Asteroids by Polyhedral Abrasion. *The Astrophysical Journal Letters, 699*(1), L13-16. doi:10.1088/0004-637X/699/1/L13

Dones, L., Brasser, R., Kaib, N., & Rickman, H. (2015). Origin and Evolution of the Cometary Reservoirs. *Space Science Reviews, 197*, 191-269. doi:10.1007/s11214-015-0223-2

Dones, L., Weissman, P. R., Levison, H. F., & Duncan, M. J. (2004). Oort Cloud Formation and Dynamics. In D. Johnstone, F. C. Adams, D. N. Lin, D. A. Neufeld, & E. Ostriker (Ed.), *Star Formation in the Interstellar Medium: In Honor of David Hollenbach, Chris McKee and Frank Shu. 323*, p. 371. San Francisco: Astronomical Society of the Pacific.

Engelhardt, T., Jedicke, R., Vereš, P., Fitzsimmons, A., Denneau, L., Beshore, E., & Meinke, B. (2017). An Observational Upper Limit on the Interstellar Number Density of Asteroids and Comets. *The Astronomical Journal, 153*(3), 133. doi:10.3847/1538-3881/aa5c8a

Feng, F., & Jones, H. R. (2018). 'Oumuamua as a Messenger from the Local Association. *The Astrophysical Journal Letters, 852*(2), L27. doi:10.3847/2041-8213/aaa404

Fitzsimmons, A., Hainaut, O., Meech, K. J., Jehin, E., Moulane, Y., Opitom, C., . . . Snodgrass, C. (2019). Detection of CN Gas in Interstellar Object 2I/Borisov. *The Astrophysical Journal Letters, 885*(1), L9. doi:10.3847/2041-8213/ab49fc

Fraser, W. C., Brown, M. E., Morbidelli, A., Parker, A., & Batygin, K. (2014). The Absolute Magnitude Distribution of Kuiper Belt Objects. *The Astrophysical Journal, 782*(2), 100. doi:10.1088/0004-637X/782/2/100

Gaudi, B. S., Bennett, D. P., Udalski, A., Gould, A., Christie, G. W., Maoz, D., . . . et al. (2008). Discovery of a Jupiter/Saturn Analog with Gravitational Microlensing. *Science, 319*(5865), 927-. doi:10.1126/science.1151947

Grimm, S. L., Demory, B.-O., Gillon, M., Dorn, C., Agol, E., Burdanov, A., . . . et al. (2018). The nature of the TRAPPIST-1 exoplanets. *Astronomy & Astrophysics, 613*, A68. doi:10.1051/0004-6361/201732233

Hallatt, T., & Wiegert, P. (2020). The Dynamics of Interstellar Asteroids and Comets within the Galaxy: An Assessment of Local Candidate Source Regions for 1I/'Oumuamua and 2I/Borisov. *The Astronomical Journal, 159*(4), 147. doi:10.3847/1538-3881/ab7336

Harrison, J. H., Bonsor, A., & Madhusudhan, N. (2018). Polluted white dwarfs: constraints on the origin and geology of exoplanetary material. *Monthly Notices of the Royal Astronomical Society, 479*(3), 3814-3841. doi:10.1093/mnras/sty1700

Hayashi, C. (1981). Structure of the Solar Nebula, Growth and Decay of Magnetic Fields and Effects of Magnetic and Turbulent Viscosities on the Nebula. *Progress of Theoretical Physics Supplement, 70*, 35-53. doi:10.1143/PTPS.70.35

Holman, M. J., & Wiegert, P. A. (1999). Long-Term Stability of Planets in Binary Systems. *The Astronomical Journal, 117*(1), 621-628. doi:10.1086/300695

Hopkins, P. F. (2016). Jumping the gap: the formation conditions and mass function of `pebble-pile' planetesimals. *Monthly Notices of the Royal Astronomical Society, 456*(3), 2383-2405. doi:10.1093/mnras/stv2820

Housen, K. R., & Holsapple, K. A. (2011). Ejecta from impact craters. *Icarus, 211*(1), 856-875. doi:10.1016/j.icarus.2010.09.017

Hyodo, R., & Genda, H. (2020). Escape and Accretion by Cratering Impacts: Formulation of Scaling Relations for High-speed Ejecta. *The Astrophysical Journal, 898*(1), 30. doi:10.3847/1538-4357/ab9897

Jewitt, D. (2003). Project Pan-STARRS and the Outer Solar System. *Earth, Moon, and Planets, 92*(1), 465-476. doi:10.1023/B:MOON.0000031961.88202.60

Jewitt, D., Hui, M.-T., Kim, Y., Mutchler, M., Weaver, H., & Agarwal, J. (2020). The Nucleus of Interstellar Comet 2I/Borisov. *The Astrophysical Journal Letters, 888*(2), L23. doi:10.3847/2041-8213/ab621b

Jewitt, D., Luu, J., Rajagopal, J., Kotulla, R., Ridgway, S., Liu, W., & Augusteijn, T. (2017). Interstellar Interloper 1I/2017 U1: Observations from the NOT and WIYN Telescopes. *The Astrophysical Journal Letters, 850*(2), L36. doi:10.3847/2041-8213/aa9b2f

Kaib, N. A., & Quinn, T. (2008). The formation of the Oort cloud in open cluster environments. *Icarus, 197*(1), 221-238. doi:10.1016/j.icarus.2008.03.020

Kaib, N. A., & Quinn, T. (2009). Reassessing the Source of Long-Period Comets. *Science, 325*(5945), 1234-. doi:10.1126/science.1172676

Kimura, J., & Kamata, S. (2020). Stability of the subsurface ocean of Pluto. *Planetary and Space Science, 181*, 104828. doi:10.1016/j.pss.2019.104828

Kiss, C., Marton, G., Parker, A. H., Grundy, W. M., Farkas-Takács, A., Stansberry, J., . . . Vinkó, J. (2019). The mass and density of the dwarf planet (225088) 2007 OR10. *Icarus, 334*, 3-10. doi:10.1016/j.icarus.2019.03.013

Kraus, R. G., Senft, L. E., & Stewart, S. T. (2011). Impacts onto H2O ice: Scaling laws for melting, vaporization, excavation, and final crater size. *Icarus, 214*(2), 724-738. doi:10.1016/j.icarus.2011.05.016

Kroupa, P. (2001). On the variation of the initial mass function. *Monthly Notices of the Royal Astronomical Society, 322*(2), 231-246. doi:10.1046/j.1365-8711.2001.04022.x

Laskar, J. (1996). Large Scale Chaos and Marginal Stability in the Solar System. *Celestial Mechanics and Dynamical Astronomy, 64*(1-2), 115-162. doi:10.1007/BF00051610

Lin, D. N., & Ida, S. (1997). On the Origin of Massive Eccentric Planets. *The Astrophysical Journal, 477*(2), 781-791. doi:10.1086/303738

Lisse, C. M., Jackson, A. P., Wolk, S. J., Snios, B. T., Desch, S. J., Unterborn, C., . . . Panic, O. (2019). M-stars Are Fast and Neat and A-stars Are Slow and Messy at Late-stage Rocky Planet Formation. *Research Notes of the American Astronomical Society, 3*(7), 90. doi:10.3847/2515-5172/ab2e0e

Lodders, K. (2003). Solar System Abundances and Condensation Temperatures of the Elements. *The Astrophysical Journal, 591*(2), 1220-1247. doi:10.1086/375492

Lorenzi, V., Pinilla-Alonso, N., & Licandro, J. (2015). Rotationally resolved spectroscopy of dwarf planet (136472) Makemake. *Astronomy & Astrophysics, 577*, A86. doi:10.1051/0004-6361/201425575

Malhotra, R. (1993). The origin of Pluto's peculiar orbit. *Nature, 365*(6449), 819-821. doi:10.1038/365819a0

McKinnon, W. B., Glein, C. R., Bertrand, T., & Rhoden, A. (2020). Formation, compositon and history of the Pluto system: A post-New Horizons synthesis. In S. Stern, & e. al (Eds.), *The Pluto system after New Horizons.* in press.

McKinnon, W. B., Nimmo, F., Wong, T., Schenk, P. M., White, O. L., Roberts, J. H., . . . et al. (2016). Convection in a volatile nitrogen-ice-rich layer drives Pluto's geological vigour. *Nature, 534*(7605), 82-85. doi:10.1038/nature18289

McKinnon, W. B., Schenk, P. M., Mao, X., Moore, J. M., Spencer, J. R., Nimmo, F., . . . Team, N. H. (2017). Impact Origin of Sputnik Planitia Basin, Pluto. *48th Lunar and Planetary Science Conference* (p. 2854). The Woodlands, TX: Lunar and Planetary Science Institute.

Meech, K. J., Weryk, R., Micheli, M., Kleyna, J. T., Hainaut, O. R., Jedicke, R., . . . Fle. (2017). A brief visit from a red and extremely elongated interstellar asteroid. *Nature, 552*(7685), 378-381. doi:10.1038/nature25020

Melosh, H. J. (1989). *Impact cratering : a geologic process.* Oxford, UK: Oxford University Press.

Micheli, M., Farnocchia, D., Meech, K. J., Buie, M. W., Hainaut, O. R., Prialnik, D., . . . Petropoulos, A. E. (2018). Non-gravitational acceleration in the trajectory of 1I/2017 U1 ('Oumuamua). *Nature, 559*, 223-226. doi:10.1038/s41586-018-0254-4

Mor, R., Robin, A. C., Figueras, F., Roca-Fàbrega, S., & Luri, X. (2019). Gaia DR2 reveals a star formation burst in the disc 2-3 Gyr ago. *Astronomy & Astrophysics, 624*, L1. doi:10.1051/0004-6361/201935105

Morbidelli, A., & Raymond, S. N. (2016). Challenges in planet formation. *Journal of Geophysical Research: Planets, 121*(10), 1962-1980. doi:10.1002/2016JE005088

Morbidelli, A., & Rickman, H. (2015). Comets as collisional fragments of a primordial planetesimal disk. *Astronomy & Astrophysics, 583*, A43. doi:10.1051/0004-6361/201526116

Moro-Martín, A., Turner, E. L., & Loeb, A. (2009). Will the Large Synoptic Survey Telescope Detect Extra-Solar Planetesimals Entering the Solar System? *The Astrophysical Journal, 704*(1), 733-704. doi:10.1088/0004-637X/704/1/733

Nesvorný, D., & Vokrouhlický, D. (2016). Neptune's Orbital Migration Was Grainy, Not Smooth. *The Astrophysical Journal, 825*(2), 94. doi:10.3847/0004-637X/825/2/94

Nesvorný, D., Vokrouhlický, D., Bottke, W. F., & Levison, H. F. (2018). Evidence for very early migration of the Solar System planets from the Patroclus-Menoetius binary Jupiter Trojan. *Nature Astronomy, 2*, 878-882. doi:10.1038/s41550-018-0564-3

Neveu, M., Desch, S. J., Shock, E. L., & Glein, C. R. (2015). Prerequisites for explosive cryovolcanism on dwarf planet-class Kuiper belt objects. *Icarus, 246*, 48-64. doi:10.1016/j.icarus.2014.03.043

Opitom, C., Fitzsimmons, A., Jehin, E., Moulane, Y., Hainaut, O., Meech, K. J., . . . Kleyna, J. T. (2019). 2I/Borisov: A C2-depleted interstellar comet. *Astronomy & Astrophysics, 631*, L8. doi:10.1051/0004-6361/201936959

'Oumuamua ISSI Team, Bannister, M. T., Bhandare, A., Dybczyński, P. A., Fitzsimmons, A., Guilbert-Lepoutre, A., . . . Ye, Q. (2019). The natural history of `Oumuamua. *Nature Astronomy, 3*, 594-602. doi:10.1038/s41550-019-0816-x

Owen, T. C., Roush, T. L., Cruikshank, D. P., Elliot, J. L., Young, L. A., de Bergh, C., . . . Bartholomew, M. J. (1993). Surface Ices and the Atmospheric Composition of Pluto. *Science, 261*(5122), 745-748. doi:10.1126/science.261.5122.745

Pitjeva, E. V., & Pitjev, N. P. (2018). Mass of the Kuiper belt. *Celestial Mechanics and Dynamical Astronomy, 130*(9), 57. doi:10.1007/s10569-018-9853-5

Poch, O., Istiqomah, I., Quirico, E., Beck, P., Schmitt, B., Theulé, P., . . . et al. (2020). Ammonium salts are a reservoir of nitrogen on a cometary nucleus and possibly on some asteroids. *Science, 367*(6483), aaw7462. doi:10.1126/science.aaw7462

Portegies Zwart, S., Torres, S., Pelupessy, I., Bédorf, J., & Cai, M. X. (2018). The origin of interstellar asteroidal objects like 1I/2017 U1 `Oumuamua. *Monthly Notices of the Royal Astronomical Society: Letters, 479*(1), L17-22. doi:10.1093/mnrasl/sly088

Protopapa, S., Grundy, W. M., Reuter, D. C., Hamilton, D. P., Dalle Ore, C. M., Cook, J. C., . . . et al. (2017). Pluto's global surface composition through pixel-by-pixel Hapke modeling of New Horizons Ralph/LEISA data. *Icarus, 287*, 218-228. doi:10.1016/j.icarus.2016.11.028

Rasio, F. A., & Ford, E. B. (1996). Dynamical instabilities and the formation of extrasolar planetary systems. *Science, 274*, 954-956. doi:10.1126/science.274.5289.954

Raymond, S. N., Armitage, P. J., & Gorelick, N. (2010). Planet-Planet Scattering in Planetesimal Disks. II. Predictions for Outer Extrasolar Planetary Systems. *The Astrophysical Journal, 711*(2), 772-795. doi:10.1088/0004-637X/711/2/772

Raymond, S. N., Armitage, P. J., & Veras, D. (2018). Interstellar Object 'Oumuamua as an Extinct Fragment of an Ejected Cometary Planetesimal. *The Astrophysical Journal Letters, 856*(1), L7. doi:10.3847/2041-8213/aab4f6

Raymond, S. N., Armitage, P. J., Moro-Martín, A., Booth, M., Wyatt, M. C., Armstrong, J. C., . . . West, A. A. (2011). Debris disks as signposts of terrestrial planet formation. *Astronomy & Astrophysics, 530*, A62. doi:10.1051/0004-6361/201116456

Raymond, S. N., Armitage, P. J., Moro-Martín, A., Booth, M., Wyatt, M. C., Armstrong, J. C., . . . West, A. A. (2012). Debris disks as signposts of terrestrial planet formation. II. Dependence of exoplanet architectures on giant planet and disk properties. *Astronomy & Astrophysics, 541*, A11. doi:10.1051/0004-6361/201117049

Raymond, S. N., Armitage, P. J., Veras, D., Quintana, E. V., & Barclay, T. (2018). Implications of the interstellar object 1I/'Oumuamua for planetary dynamics and planetesimal formation. *Monthly Notices of the Royal Astronomical Society, 476*(3), 3031-3038. doi:10.1093/mnras/sty468

Robbins, S. J., & Hynek, B. M. (2011). Secondary crater fields from 24 large primary craters on Mars: Insights into nearby secondary crater production. *Journal of Geophysical Research, 116*(E10), E10003. doi:10.1029/2011JE003820

Robuchon, G., & Nimmo, F. (2011). Thermal evolution of Pluto and implications for surface tectonics and a subsurface ocean. *Icarus, 216*(2), 426-439. doi:10.1016/j.icarus.2011.08.015

Rubin, M., Altwegg, K., Balsiger, H., Bar-Nun, A., Berthelier, J. -J., Bieler, A., . . . et al. (2015). Molecular nitrogen in comet 67P/Churyumov-Gerasimenko indicates a low formation temperature. *Science, 348*(6231), 232-235. doi:10.1126/science.aaa6100

Schaller, E. L., & Brown, M. E. (2007). Volatile Loss and Retention on Kuiper Belt Objects. *The Astrophysical Journal, 659*(1), L61-64. doi:10.1086/516709

Seager, S. (2014). The future of spectroscopic life detection on exoplanets. *Proceedings of the National Academy of Sciences, 111*(35), 12634-12640. doi:10.1073/pnas.1304213111

Shannon, A., & Dawson, R. (2018). Limits on the number of primordial Scattered disc objects at Pluto mass and higher from the absence of their dynamical signatures on the present-day trans-Neptunian Populations. *Monthly Notices of the Royal Astronomical Society, 480*(2), 1870-1882. doi:10.1093/mnras/sty1930

Shannon, A., Jackson, A. P., & Wyatt, M. C. (2019). Oort cloud asteroids: collisional evolution, the Nice Model, and the Grand Tack. *Monthly Notices of the Royal Astronomical Society, 485*(4), 5511-5518. doi:10.1093/mnras/stz776

Shannon, A., Jackson, A. P., Veras, D., & Wyatt, M. (2015). Eight billion asteroids in the Oort cloud. *Monthly Notices of the Royal Astronomical Society, 446*(2), 2059-2064. doi:10.1093/mnras/stu2267

Singer, K. N., McKinnon, W. B., Gladman, B., Greenstreet, S., Bierhaus, E. B., Stern, S. A., . . . et al. (2019). Impact craters on Pluto and Charon indicate a deficit of small Kuiper belt objects. *Science, 363*(6430), 955-959. doi:10.1126/science.aap8628

Swan, A., Farihi, J., Koester, D., Hollands, M., Parsons, S., Cauley, P. W., . . . Gänsicke, B. T. (2019). Interpretation and diversity of exoplanetary material orbiting white dwarfs. *Monthly Notices of the Royal Astronomical Society, 490*(1), 202-218. doi:10.1093/mnras/stz2337

Tegler, S. C., Cornelison, D. M., Grundy, W. M., Romanishin, W., Abernathy, M. R., Bovyn, M. J., . . . Vilas, F. (2010). Methane and Nitrogen Abundances on Pluto and Eris. *The Astrophysical Journal, 725*(1), 1296-1305. doi:10.1088/0004-637X/725/1/1296

Tegler, S. C., Grundy, W. M., Vilas, F., Romanishin, W., Cornelison, D. M., & Consolmagno, G. J. (2008). Evidence of N2-ice on the surface of the icy dwarf Planet 136472 (2005 FY9). *Icarus, 195*(2), 844-850. doi:10.1016/j.icarus.2007.12.015

Trilling, D. E., Mommert, M., Hora, J. L., Farnocchia, D., Chodas, P., Giorgini, J., . . . et al. (2018). Spitzer Observations of Interstellar Object 1I/'Oumuamua. *The Astronomical Journal, 156*(6), 261. doi:10.3847/1538-3881/aae88f

Tsiganis, K., Gomes, R., Morbidelli, A., & Levison, H. F. (2005). Origin of the orbital architecture of the giant planets of the Solar System. *Nature, 435*(7041), 459-461. doi:10.1038/nature03539

Vokrouhlický, D., Nesvorný, D., & Dones, L. (2019). Origin and Evolution of Long-period Comets. *The Astronomical Journal, 157*(5), 181. doi:10.3847/1538-3881/ab13aa

Volk, K., & Gladman, B. (2015). Consolidating and Crushing Exoplanets: Did It Happen Here? *The Astrophysical Journal Letters, 806*(2), L26. doi:10.1088/2041-8205/806/2/L26

Weidenschilling, S. J. (1977). The Distribution of Mass in the Planetary System and Solar Nebula. *Astrophysics & Space Science, 51*(1), 153-158. doi:10.1007/BF00642464

Weidenschilling, S. J., & Marzari, F. (1996). Gravitational scattering as a possible origin for giant planets at small stellar distances. *Nature, 384*(6610), 619-621. doi:10.1038/384619a0

Widmark, A., & Monari, G. (2019). The dynamical matter density in the solar neighbourhood inferred from Gaia DR1. *Monthly Notices of the Royal Astronomical Society, 482*(1), 262-277. doi:10.1093/mnras/sty2400

Wittenmyer, R. A., Wang, S., Horner, J., Butler, R. P., Tinney, C. G., Carter, B. D., . . . Johns, D. (2020). Cool Jupiters greatly outnumber their toasty siblings: occurrence rates from the Anglo-Australian Planet Search. *Monthly Notices of the Royal Astronomical Society, 492*(1), 377-383. doi:10.1093/mnras/stz3436

Wyatt, M. C., Bonsor, A., Jackson, A. P., Marino, S., & Shannon, A. (2017). How to design a planetary system for different scattering outcomes: giant impact sweet spot, maximizing exocomets, scattered discs. *Monthly Notices of the Royal Astronomical Society, 464*(3), 3385-3407. doi:10.1093/mnras/stw2633